\documentclass[preprint,aps,showpacs,pre,amsmath,amssymb]{revtex4-1}
\usepackage{amssymb}
\usepackage{physics}
\usepackage{xcolor}

\usepackage{graphicx}
\usepackage{xcolor}
\usepackage{soul}  

\usepackage{booktabs}
\usepackage{multirow}

\usepackage{tikz}

\usepackage{dsfont}
\usepackage{appendix}
\usepackage{enumitem}
\usepackage{booktabs}
\usepackage{soul}

\usepackage{float}
\usepackage{placeins}
\usepackage{capt-of}

\usetikzlibrary{positioning, arrows.meta, shapes.geometric, backgrounds, fit}

\begin{document}

\title{Fine-Tuned Machine-Learned Interatomic Potentials for Structural and Vibrational Properties of Twisted 2D Materials}
	
\author{Viet-Anh Tran\textsuperscript{1}}
\email{anh.tran@uclouvain.be, $^\dagger$viet-hung.nguyen@uclouvain.be, $^\ddagger$wei.chen@uclouvain.be, \textcolor{white}{a a a a a a a a aaa} $^\S$gian-marco.rignanese@uclouvain.be, $^\P$jean-christophe.charlier@uclouvain.be}
\author{Viet-Hung Nguyen\textsuperscript{1}$^{,\dagger}$}
\author{Wei Chen\textsuperscript{1}$^{,\ddagger}$}
\author{Gian-Marco Rignanese\textsuperscript{1,2}$^{,\S}$}
\author{Jean-Christophe Charlier\textsuperscript{1}$^{,\P}$}
\affiliation{$^1$Institute of Condensed Matter and Nanosciences, 
Universit\'{e} catholique de Louvain (UCLouvain), Chemin 
des \'etoiles 8, B-1348 Louvain-la-Neuve, Belgium}
\affiliation{$^2$WEL Research Institute, Avenue Pasteur, 6, B-1300 Wavre, Belgium}

\begin{abstract} 
Twisted van der Waals bilayers form moir\'e superlattices whose structural and vibrational properties are highly sensitive to variations in local stacking registry and the degree of atomic reconstruction, yet accurate atomistic modeling of these systems at the DFT level remains computationally prohibitive at small twist angles. We investigate machine-learned interatomic potentials for moir\'e systems, using twisted bilayer graphene, \textit{h}-BN, and MoS$_2$ as representative materials spanning a broad spectrum of mechanical compliance and atomic reconstruction behavior. We show that fine-tuning universal atomistic foundation models is essential to achieve DFT accuracy for layered materials, as broadly trained foundation models prove insufficient for resolving the subtle interlayer energetics that govern atomic reconstruction. Through local strain tensor analysis and the phonon band unfolding technique, our fine-tuned MACE model reveals a consistent reconstruction-induced strain landscape in all three materials, with extended low-energy stacking domains separated by narrow soliton lines where deformation concentrates. The system progressively optimizes the local stacking registry within each domain, giving rise to a spatially structured deformation field whose amplitude scales with the mechanical compliance of the material and can be further tuned by external perturbation. The obtained results of both atomic reconstructed structures and moiré phonon spectra present a good agreement with the reported experiments, thereby demonstrating the accuracy and efficiency of our methodology in modeling of these large scale nanomaterials.

%This deformation field leaves distinct signatures in the phonon spectrum that are in good agreement with experimental Raman observations, a level of fidelity that simple classical force-field potentials cannot achieve owing to their systematic overestimation of the strain by a factor of several. 
\end{abstract}

\maketitle

\section{Introduction}

Two-dimensional (2D) materials such as graphene, hexagonal boron nitride (\textit{h}-BN), and transition-metal dichalcogenides (TMDs) exhibit a remarkable range of electronic, mechanical, and optical properties that are fundamentally distinct from those of their bulk forms, arising from reduced dimensionality of their crystalline lattice~\cite{novoselov2004electric,geim2013van,manzeli20172d}. Assembling these atomically thin crystals into van der Waals heterostructures, held together by weak interlayer forces without the constraints of lattice matching, provides a versatile route to engineer material properties beyond what any single constituent can offer~\cite{geim2013van,novoselov20162d}. Among the available control parameters in such heterostructures, the relative twist angle between adjacent layers has emerged as a particularly powerful and experimentally accessible knob, enabling systematic tuning of interlayer coupling, stacking configuration, and lattice reconstruction~\cite{carr2018relaxation,weston2020reconstruction}. Stacking two-dimensional materials with a relative twist angle or lattice mismatch generates moir\'e superlattices whose long-wavelength modulation can strongly reshape the low-energy electronic, vibrational, and structural properties of the constituent layers~\cite{bistritzer2011moire, nuckolls2024microscopic,cao2018correlated, cao2018unconventional,khalaf2019magic, devakul2021magic,zhang20214}. This capability has driven the rapid development of twistronics, with seminal discoveries ranging from correlated insulating and superconducting phases in magic-angle twisted bilayer graphene~\cite{cao2018correlated, cao2018unconventional} to correlation-driven and fractional states in twisted TMD moir\'e systems~\cite{wang2020correlated,cai2023signatures}. Variation in local stacking registry, out-of-plane corrugation, and local strain can substantially modify the electronic structure and even alter the dominant reconstruction pattern~\cite{yoo2019atomic, weston2020reconstruction,kazmierczak2021strain}. This sensitivity creates a severe computational challenge, as the relevant energy differences between competing local registries are often only a few meV per atom~\cite{siddiqui2025understanding,carr2018relaxation}, whereas the corresponding moir\'e supercells generally contain a large number of atoms. Experiments have directly visualized this structural complexity through atomic-resolution imaging of reconstruction patterns in van der Waals bilayers~\cite{yoo2019atomic,weston2020reconstruction}, as well as vibrational probes that reveal twist-dependent phonon renormalization in reconstructed moir\'e superlattices~\cite{quan2021phonon,gadelha2021localization,solanki2025anomalies}. Because the emergent phenomena, such as flat bands, strong correlation effects, and localized states, depend sensitively on these atomic details, accurate theoretical modeling must capture the full spatial distribution of atomic displacements, interlayer corrugation, and in-plane deformation.

Density functional theory (DFT) provides the benchmark accuracy needed to resolve these subtle energetic differences, but its computational cost scales unfavorably with system size and rapidly becomes prohibitive at moir\'e length scales~\cite{carr2020electronic}. For twist angles below a few degrees, precisely where structural reconstruction becomes strongest and the most interesting physics often emerges, even a single structural relaxation may require supercells containing several thousand atoms. Comprehensive studies of twist-angle dependence, structural or vibrational properties are therefore difficult to carry out entirely at the DFT level. This computational barrier requires an alternative modeling strategy. Classical interatomic potentials enable large-scale atomistic simulations, and specialized force fields have been developed for specific layered systems, pairing intralayer bonding descriptions (Stillinger--Weber~\cite{jiang2013stillinger, hossain2018stillinger}, REBO~\cite{brenner2002second}, or Tersoff~\cite{tersoff1988empirical}) with interlayer coupling terms (such as the Kolmogorov--Crespi or DRIP potentials)~\cite{kolmogorov2005registry, naik2019kolmogorov, wen2018dihedral}. However, as we demonstrate below, classical force fields are insufficient to accurately describe the local structural deformation as well as the vibrational properties of reconstructed moir\'e superlattices, a limitation that originates from the rigidity of their prescribed functional forms in capturing stacking-dependent interlayer energetics. Machine-learned interatomic potentials (MLIPs) thus provide an attractive middle ground to overcome this limitation by learning energies and forces directly from first-principles data while retaining computational efficiency much closer to that of classical force fields. Recent advances in equivariant neural-network architectures, including Allegro~\cite{musaelian2023learning}, NequIP~\cite{batzner2022nequip}, and MACE~\cite{batatia2022mace,batatia2025foundation}, have substantially improved the accuracy and data efficiency of atomistic machine learning for materials problems in which many-body geometric effects are important. More recently, the field has moved toward foundation models trained on broad and chemically diverse datasets, such as CHGNet~\cite{deng2023chgnet} and MACE~\cite{batatia2025foundation}, designed to generalize across large regions of chemical space without requiring system-specific training from scratch. Their emergence raises a natural question for moir\'e materials as to whether a broadly trained foundation model could already predict the meV accuracy needed for twisted layered materials relaxation, or if additional fine-tuning remains indispensable.

This question is especially pressing for van der Waals bilayers, which serve as the prototypical platform for studying moiré physics. Unlike general pristine systems, twisted bilayers explore a highly constrained configuration space characterized by periodic variations around high-symmetry stacking configurations, with coupled in-plane deformation and out-of-plane corrugation. Foundation models trained on chemically broad databases may reproduce the overall chemistry of the constituent materials while still missing the delicate interlayer energy balance that controls moir\'e reconstruction. MLIPs trained for computing layered systems have been also recently reported ~\cite{siddiqui2025understanding,siddiqui2024machine}, but a thorough comparison between foundation and fine-tuned models, together with the physical insights that such accuracy makes accessible, remains limited. While ad hoc MLIPs trained from scratch for specific layered systems have recently been reported to address this accuracy gap~\cite{siddiqui2025understanding,siddiqui2024machine}, fine-tuning a pretrained foundation model offers a more data-efficient alternative by building on the broad chemical knowledge already encoded in the foundation model, yet this route remains unexplored.

In this work, we investigate MLIPs for moir\'e systems and demonstrate that additional fine-tuning is essential for reliable predictions, a point established through direct comparison with foundation models across twisted bilayer graphene, \textit{h}-BN, and MoS$_2$, three materials that span a broad spectrum of electronic character, elastic properties, and interlayer energy corrugation, and thereby cover a wide range of reconstruction behavior. Beyond model development, we apply these high-fidelity potentials to investigate the evolution of reconstruction-induced strain and its impact on the vibrational properties of moiré superlattices. Our results reveal a consistent strain landscape in the low-angle regime and provide quantitative agreement with experimental phonon renormalization data, demonstrating that these models can capture the delicate physics of reconstructed van der Waals interfaces.

\section{Methodology}

\subsection{Fine-tuning Procedure}

\begin{figure}[htbp]
\centering
\includegraphics[width=0.75\textwidth]{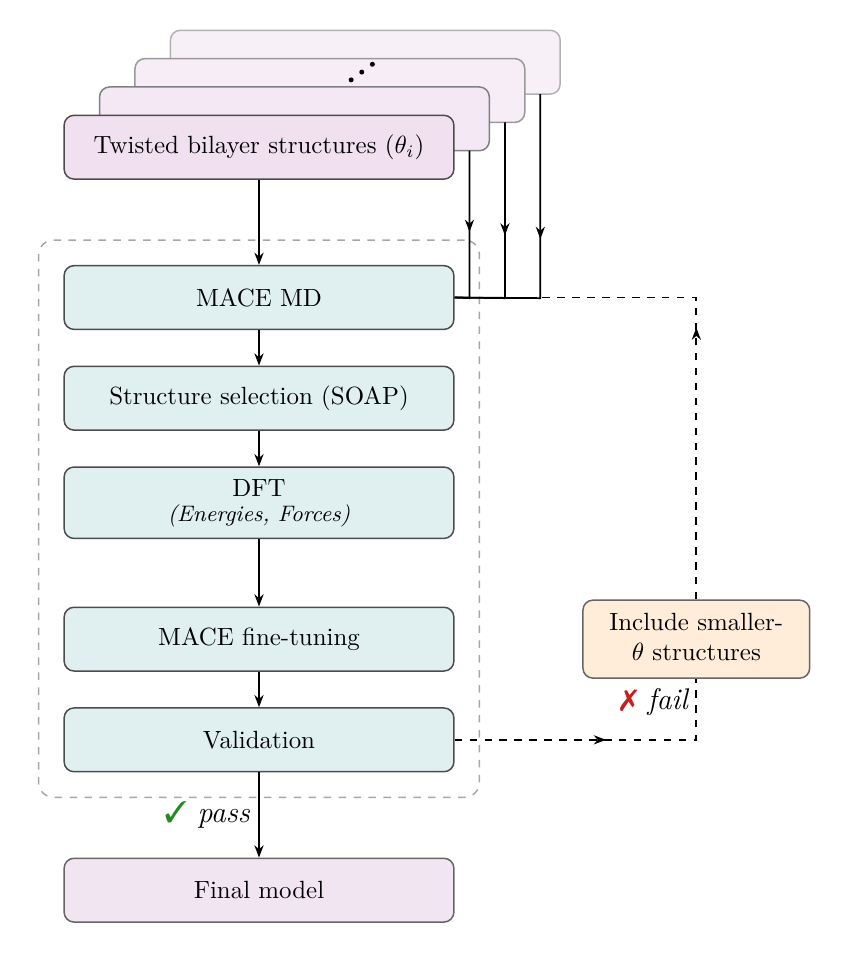}
\caption{Schematic fine-tuning workflow for twisted bilayer systems. Finite-temperature molecular dynamics trajectories generated with the MACE foundation model are subsampled using SOAP descriptors, labeled with DFT calculations (energies and forces), and used to fine-tune the foundation MACE model.}
\label{fig:fig01}
\end{figure}

Despite the impressive transferability of foundation models trained on large, chemically diverse datasets~\cite{batatia2025foundation}, they struggle to capture the highly specific energetics of twisted bilayer systems, where structural relaxation is driven by subtle energy differences as small as a few meV per atom~\cite{mostaani2015quantum,gilbert2019alternative,arnold2023relaxation}. Our fine-tuning dataset is designed specifically around this requirement. For each material, thermally distorted configurations are drawn from molecular dynamics (MD) trajectories run at 300~K using the MACE-OMAT foundation model, trained on the Open Materials 2024 (OMat24) dataset of inorganic crystals~\cite{barroso2024open}. This model is known to have a very good accuracy for structural and phonon properties, both of which are essential for reliably describing the reconstruction behavior and lattice dynamics of twisted van der Waals bilayers. The above distorted configurations are supplemented by structures spanning the relevant high-symmetry registries and intermediate translations. These configurations, which are detailed in Table~\ref{tab:registry_coverage}, are essential for the fine-tuning process because they represent the fundamental local environments that define the moir\'e energy landscape. Including these specific registries provides the model with the necessary information to resolve the subtle interlayer energy differences that dictate the competition between domain formation and lattice strain during structural reconstruction. 

The overall fine-tuning strategy is illustrated in Figure~\ref{fig:fig01}. Starting from large, computationally tractable twist angles such as $21.79^\circ$, the pool is progressively extended to smaller angles, reaching around $3.15^\circ$, to guarantee adequate coverage of the reconstructed regime. Representative configurations (20--30 per twist angle) are selected to ensure structural diversity across the trajectory snapshots, which is quantified using Smooth Overlap of Atomic Positions (SOAP) descriptors~\cite{bartok2013representing}. The resulting database contains on the order of 150--200 configurations per material system, covering the essential local environments that govern domain formation, corrugation, and soliton-wall energetics. The energy and force labels for all selected structures are obtained by single-point DFT calculations performed with VASP~\cite{kresse1996efficient}, using the PBE functional with D3 dispersion correction~\cite{grimme2010consistent}. The MACE architecture~\cite{batatia2022mace}, initialized from the MACE-OMAT pretrained foundation model~\cite{shuaibi2024omat24}, is fine-tuned by minimizing a combined energy--force loss, 
\begin{equation} \mathcal{L} = \lambda_E \mathcal{L}_E + \lambda_F \mathcal{L}_F, 
\end{equation} 
with a weighting ratio $\lambda_E : \lambda_F = 1:10$ to prioritize force accuracy, which is the primary determinant of relaxed structural quality. The atomic environment cutoff is set to 6~\AA, as hardcoded in the MACE-OMAT foundation model; with two message-passing layers, this yields an effective interaction range of 12~\AA, sufficient to capture the relevant interlayer interactions in van der Waals bilayers. The model is fine-tuned and evaluated using an 80:20 train--test split throughout. Its performance is assessed through parity plots on the held-out test set and by direct comparison of relaxed structural properties, including interlayer spacing, stacking-dependent energetics, and the relative stability of high-symmetry registries. When discrepancies in structural relaxation indicate insufficient accuracy at smaller twist angles, additional configurations are incorporated and fine-tuning is repeated. In practice, this refinement is needed at most once per material, confirming that the foundation model provides a sufficiently good initialization for a modest, targeted dataset to suffice.

\subsection{Atomistic Modeling of Moir\'e Superlattices}

\begin{figure*}[t]
\centering
\includegraphics[width=0.8\textwidth]{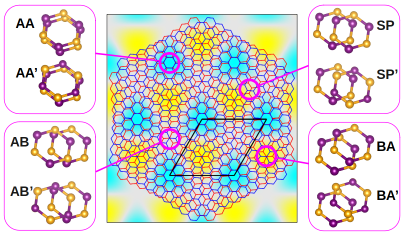}
\caption{High-symmetry stacking configurations for twisted bilayer graphene, \textit{h}-BN, and MoS$_2$. Center: representative moir\'e supercell. Surrounding panels: zoom-in views of each high-symmetry registry. In the parallel orientation ($\sim 0^\circ$), the relevant stackings are AA, AB, and BA; for graphene, AB and BA are equivalent by symmetry. In the antiparallel orientation ($\sim 60^\circ$), \textit{h}-BN and MoS$_2$ additionally exhibit AA$'$, AB$'$, and BA$'$ registries arising from the inequivalence of the two sublattice atom types.}
\label{fig:stackings}
\end{figure*}

\textbf{Geometry relaxation.} Twisted bilayer systems 
are formed by stacking two identical monolayers and 
rotating one relative to the other by a twist angle 
$\theta$. The resulting moir\'e superlattice hosts a 
spatially varying stacking registry that passes through 
a set of high-symmetry configurations depending on the 
material (Figure~\ref{fig:stackings}). At small twist 
angles, atomic reconstruction becomes significant and 
must be accounted for through structural 
relaxation~\cite{yoo2019atomic, weston2020reconstruction}. 
The energy differences between competing stacking 
registries, on the order of a few meV per 
atom~\cite{siddiqui2025understanding}, set the accuracy 
requirements for any interatomic potential used in this 
relaxation. In our study, all structural relaxations are performed 
with the Atomic Simulation Environment 
(ASE)~\cite{larsen2017atomic} until the maximum force 
component on any atom falls below $10^{-4}$~eV/\AA.

\textbf{Strain tensor analysis.} In conventional 
deformed systems such as those under uniaxial or shear 
loading, strain is spatially uniform and well described 
by simple scalar measures or global strain models. In 
twisted bilayers, however, the strain field is intrinsic 
and spatially varying across the moir\'e cell, with 
qualitatively different deformation characters in 
different stacking environments. Imposing simplified 
strain models onto such a system inevitably obscures 
this local variation. We therefore characterize the 
local deformation directly from its fundamental 
definition, computing the position-dependent strain 
tensor for each layer by comparing the local lattice 
vectors constructed from nearest-neighbor atomic 
positions with those of the ideal pristine monolayer:
\begin{equation}
\boldsymbol{\varepsilon}(\mathbf{r}) =
\begin{pmatrix}
\varepsilon_{xx}(\mathbf{r}) & \varepsilon_{xy}(\mathbf{r}) \\
\varepsilon_{yx}(\mathbf{r}) & \varepsilon_{yy}(\mathbf{r})
\end{pmatrix},
\end{equation}
where the four components $\varepsilon_{xx}(\mathbf{r})$, 
$\varepsilon_{yy}(\mathbf{r})$, $\varepsilon_{xy}(\mathbf{r})$, 
and $\varepsilon_{yx}(\mathbf{r})$ fully characterize the 
local in-plane deformation field.

\textbf{Phonon calculations.} Vibrational properties are computed from the relaxed structures obtained above using the finite-displacement method as implemented in Phonopy~\cite{togo2015firstprinciples}, with atomic displacements of 0.01~\AA. Phonon spectra are then computed using our in-house codes, which have been developed for solving the Hamiltonian/dynamical matrices of large-scale systems using optimized numerical methods, such as Green's function techniques~\cite{nguyen2023recursive} combined with other conventional diagonalization approaches. For twisted bilayer systems, the large moir\'e supercell results in extensive band folding that renders the phonon spectrum difficult to analyze and prevents direct comparison with experimental probes such as Raman spectroscopy; we therefore unfold the supercell phonon modes onto the primitive-cell Brillouin zone using the spectral-weight formalism~\cite{deretzis2014role,lamparski2020soliton,gadelha2021localization}. For a phonon mode $\lambda$ at primitive-cell wavevector $\mathbf{q}$, the spectral weight is defined as
\begin{equation}
\rho_{\lambda}(\mathbf{q})=
\frac{1}{N}
\sum_{a \in \mathrm{PC}}
\left|
\sum_{j=1}^{N}
c_{a+j}^{\lambda}\,
e^{-i\mathbf{q}\cdot\mathbf{r}_{a+j}}
\right|^{2},
\label{eq:spectral_weight}
\end{equation}
where $N$ is the number of primitive cells in the 
supercell, $a$ indexes atoms within a single primitive 
cell, $j$ labels translationally equivalent cells, 
$\mathbf{r}_{a+j}$ is the position of atom $a+j$, and 
$c_{a+j}^{\lambda}$ is the corresponding Cartesian 
component of the phonon eigenvector of mode $\lambda$. 
The sum over Cartesian directions ($x$, $y$, $z$) is 
implicit. A spectral weight close to unity indicates 
that a supercell mode retains strong primitive-cell 
Bloch character at wavevector $\mathbf{q}$; smaller 
values reflect band folding or reconstruction-induced 
mode mixing. The effect of this procedure is illustrated 
in Fig.~\ref{fig:unfoldingscheme}, where the dense 
folded phonon branches of a twisted bilayer graphene 
supercell are recovered as a clean, intensity-weighted 
primitive-cell dispersion.

\section{Results and discussions} 

\subsection{Fine-Tuning MACE for Accurate Moiré Simulations}
 
\begin{figure*}[!htbp]
\centering
\includegraphics[width=1.0\textwidth]{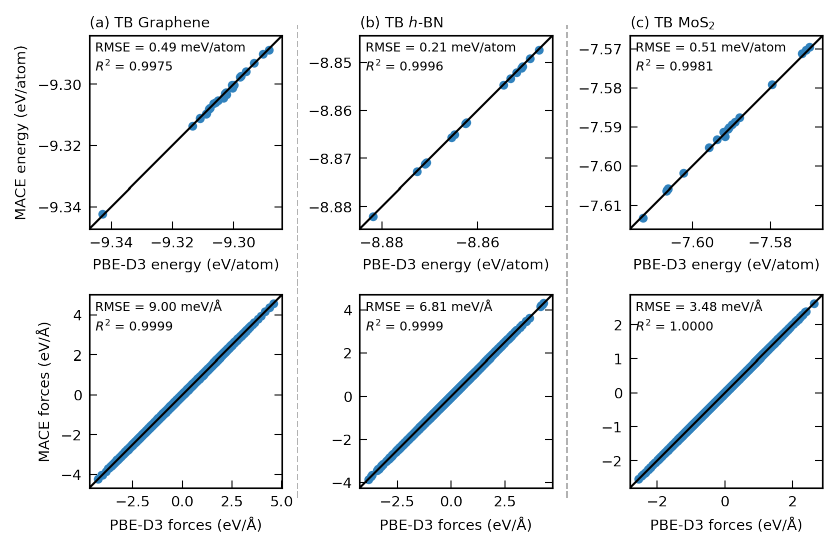}
\caption{Parity plots of fine-tuned MACE vs.\ PBE-D3 
energies (top row) and forces (bottom row) for twisted 
bilayer (TB) (a)~graphene, (b)~\textit{h}-BN, and 
(c)~MoS$_2$. The black diagonal indicates ideal 1:1 
agreement. The tight clustering of data points along 
the diagonal confirms that the fine-tuned models 
reproduce the DFT reference with no systematic bias.}
\label{fig:parity_plots}
\end{figure*}

Accurately capturing the energy landscape is the primary requirement for any reliable interatomic potential in any system. Figure~\ref{fig:parity_plots} presents the energy and force parity plots comparing our fine-tuned MACE predictions against PBE-D3 reference calculations for the three material systems studied. These results encompass a wide range of configurations from both training and test datasets, which for each material include between 150--260 representative frames (detailed registry coverage for each system is provided in Table~\ref{tab:registry_coverage}). In all three systems, data points cluster tightly along the ideal diagonal over energy ranges of several eV/atom and force magnitudes up to several eV/\AA, with no systematic bias or pronounced outliers. RMSE values reach 0.49, 0.21, and 0.51~meV/atom for energies, and 9.00, 6.81, and 3.48~meV/\AA\ for forces, in graphene, \textit{h}-BN, and MoS$_2$, respectively. Such errors are sufficiently small to resolve the subtle energy differences between competing stacking configurations that govern domain formation and reconstruction.

This energetic precision translates directly into accurate equilibrium structures, a capability that the foundation model conspicuously lacks. Table~\ref{tab:bilayer_distances} compares interlayer separations for all high-symmetry stacking configurations across the three materials, contrasting PBE-D3 reference values with predictions from both the foundation MACE (OMAT-D3) and our fine-tuned MACE. The foundation model systematically overestimates interlayer distances by up to 0.74~\AA\ for graphene, 1.05~\AA\ for \textit{h}-BN, and 0.58~\AA\ for MoS$_2$ for all stacking configurations studied, despite the inclusion of an additive D3 dispersion correction. These errors are not merely quantitative. An overestimation of this magnitude places competing stacking configurations in incorrect energetic ordering, yielding qualitatively wrong reconstruction patterns. The corresponding total energies, reported in Table~\ref{tab:bilayer_energy}, confirm that the foundation model also fails to resolve the meV-scale energy differences between stacking registries. This systematic failure can be understood from the training distribution of foundation models, which are built on chemically diverse but structurally generic datasets that do not adequately sample the narrow, registry-dependent configuration space characteristic of van der Waals bilayers. Fine-tuning on a stacking-diverse dataset built from molecular dynamics trajectories and first-principles single-point calculations directly remedies this deficiency: deviations from PBE-D3 are reduced to below 0.05~\AA\ for graphene, 0.09~\AA\ for \textit{h}-BN, and 0.04~\AA\ for MoS$_2$, for every high-symmetry configuration. This improvement is further illustrated by the interlayer distance maps of twisted bilayer systems at $\theta \approx 3.1^\circ$ (Figs.~\ref{fig:interlayer_validation_tblg} and~\ref{fig:interlayer_validation_hbn}), where the foundation model not only overestimates the interlayer separation over the entire unit cell but also fails to reproduce the spatial variation between stacking regions, yielding a qualitatively incorrect reconstruction pattern. The fine-tuned MACE, by contrast, reproduces the PBE-D3 interlayer distance map with high fidelity.

\begin{table}[!htbp]
\centering
\caption{Interlayer distances for relaxed aligned 
bilayer stackings: comparison between PBE-D3, 
foundation MACE (OMAT-D3), and fine-tuned MACE. 
Values in parentheses indicate the absolute 
deviation from the PBE-D3 reference.\\}
\label{tab:bilayer_distances}
\small
\renewcommand{\arraystretch}{0.7}
\begin{ruledtabular}
\begin{tabular}{lccc}
Bilayer system & PBE-D3 (\AA) & OMAT-D3 (\AA) & 
Fine-tuned (\AA) \\
\hline
\multicolumn{4}{l}{\textit{Graphene}} \\
~~AA stacking & 3.65 & 4.15~(0.50) & 3.60~(0.05) \\
~~AB stacking & 3.44 & 4.18~(0.74) & 3.43~(0.01) \\
\hline
\multicolumn{4}{l}{\textit{\textit{h}-BN}} \\
~~AA stacking    & 3.61 & 4.18~(0.57) & 3.57~(0.04) \\
~~AB stacking    & 3.36 & 4.36~(1.00) & 3.38~(0.02) \\
~~AA$'$ stacking & 3.37 & 4.42~(1.05) & 3.36~(0.01) \\
~~AB$'$ stacking & 3.49 & 4.30~(0.81) & 3.42~(0.07) \\
~~BA$'$ stacking & 3.65 & 4.23~(0.58) & 3.56~(0.09) \\
\hline
\multicolumn{4}{l}{\textit{MoS$_2$}} \\
~~AA stacking    & 6.66 & 7.20~(0.54) & 6.65~(0.01) \\
~~AB stacking    & 6.03 & 6.30~(0.27) & 6.06~(0.03) \\
~~AA$'$ stacking & 6.03 & 6.34~(0.31) & 6.07~(0.04) \\
~~AB$'$ stacking & 6.08 & 6.24~(0.16) & 6.04~(0.04) \\
~~BA$'$ stacking & 6.62 & 7.20~(0.58) & 6.64~(0.02) \\
\end{tabular}
\end{ruledtabular}
\end{table}

\begin{figure*}[!htbp]
\centering
\includegraphics[width=1.0\textwidth]{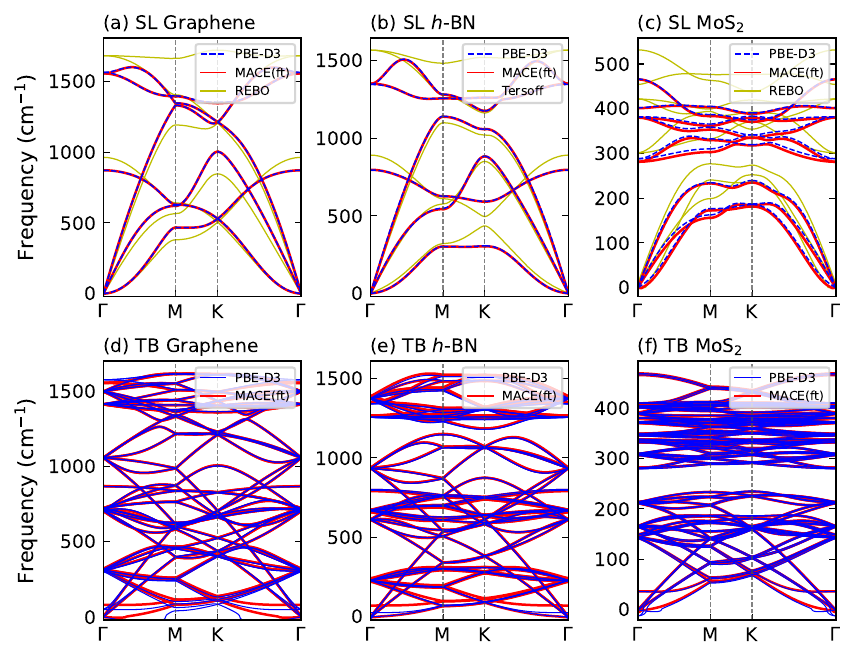}
\caption{Phonon dispersion relations for single-layer (SL) (a--c) and $\theta \approx 21^\circ$ twisted bilayer (TB) (d--f) configurations of graphene, \textit{h}-BN, and MoS$_2$. PBE-D3 (green), fine-tuned MACE (red), and classical force fields (blue; panels a--c only) are compared.}
\label{fig:phonon_comparison}
\end{figure*}

We next assess whether this accuracy extends to vibrational properties of pristine and large-angle twisted bilayer systems, where supercells remain tractable for direct DFT comparison, noting that such properties are determined by second derivatives of the potential energy surface and hence more sensitive to subtle force field errors than ground-state geometries. Figure~\ref{fig:phonon_comparison} compares phonon dispersions for single-layer (panels a--c) and $\theta \approx 21^\circ$ twisted bilayer (panels d--f) configurations of all three materials, with classical force fields included in the single-layer panels to benchmark their intrinsic accuracy against DFT prior to the twisted bilayer analysis. For single layers, fine-tuned MACE shows excellent agreement with DFT (RMSE values always below 4.1~cm$^{-1}$ for all three materials) while classical force fields show significant deviations particularly in the optical branches. For twisted bilayer configurations, where moir\'e zone folding produces hundreds of additional branches, fine-tuned MACE maintains this agreement over the full Brillouin zone, including the low-frequency interlayer shear and breathing modes that are most directly sensitive to stacking registry and interlayer coupling \cite{liang2017low,huang2016low}. Taken together, these benchmarks establish that the fine-tuned MACE models achieve near-DFT accuracy in energetics, equilibrium structure, and lattice dynamics, consistently for three chemically distinct systems, and at a level that foundation models and classical force fields cannot match. Additional validation tests, including force parity plots resolved by stacking order parameter and quasi-one-dimensional moir\'e structures following the surrogate approach of Ref.~\cite{georgaras2025accurate}, consistently confirm that the foundation model underperforms for layered materials while fine-tuned MACE maintains near-DFT accuracy (Figs.~\ref{fig:stackingparams_tblg}--\ref{fig:quasi1D}).
 
\subsection{Reconstruction-Induced Strain Landscape}
\label{sec:reconstruction}

With this accuracy established, the fine-tuned models provide reliable access to the low-angle regime where reconstruction is strong and DFT becomes prohibitively expensive. Atomic reconstruction in twisted bilayers arises from the competition between interlayer interaction energy, which drives atoms toward energetically favorable stacking registries, and intralayer elastic energy, which resists the resulting lattice deformation. The extent and spatial distribution of the resulting displacement and strain fields depend sensitively on the balance between these two energy scales and vary markedly from one material to another, as demonstrated in the following sections.

\subsubsection{Out-of-plane Moiré Corrugation and In-Plane Displacements}

\begin{figure*}[htbp]
\centering
\includegraphics[width=1.0\textwidth]{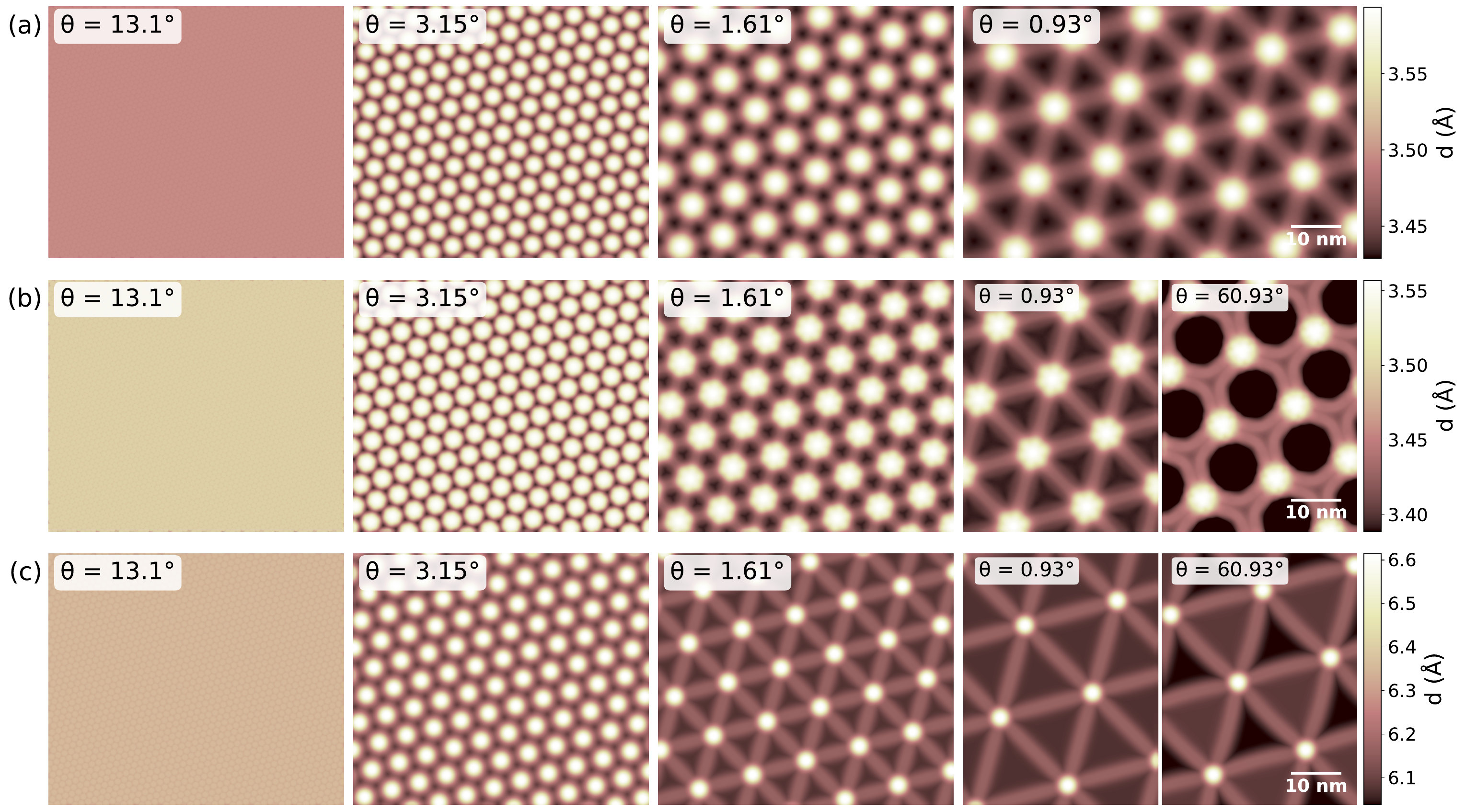}
\caption{Interlayer distance maps across the moir\'{e} unit cell for twisted bilayer
(a)~graphene, (b)~\textit{h}-BN, and (c)~MoS$_2$ at four representative twist angles. For \textit{h}-BN and MoS$_2$, an additional antiparallel-orientation panel is included as representation. Color scale indicates the local interlayer separation.}
\label{fig:interlayer}
\end{figure*}

\begin{figure*}[htbp]
\centering
\includegraphics[width=1.0\textwidth]{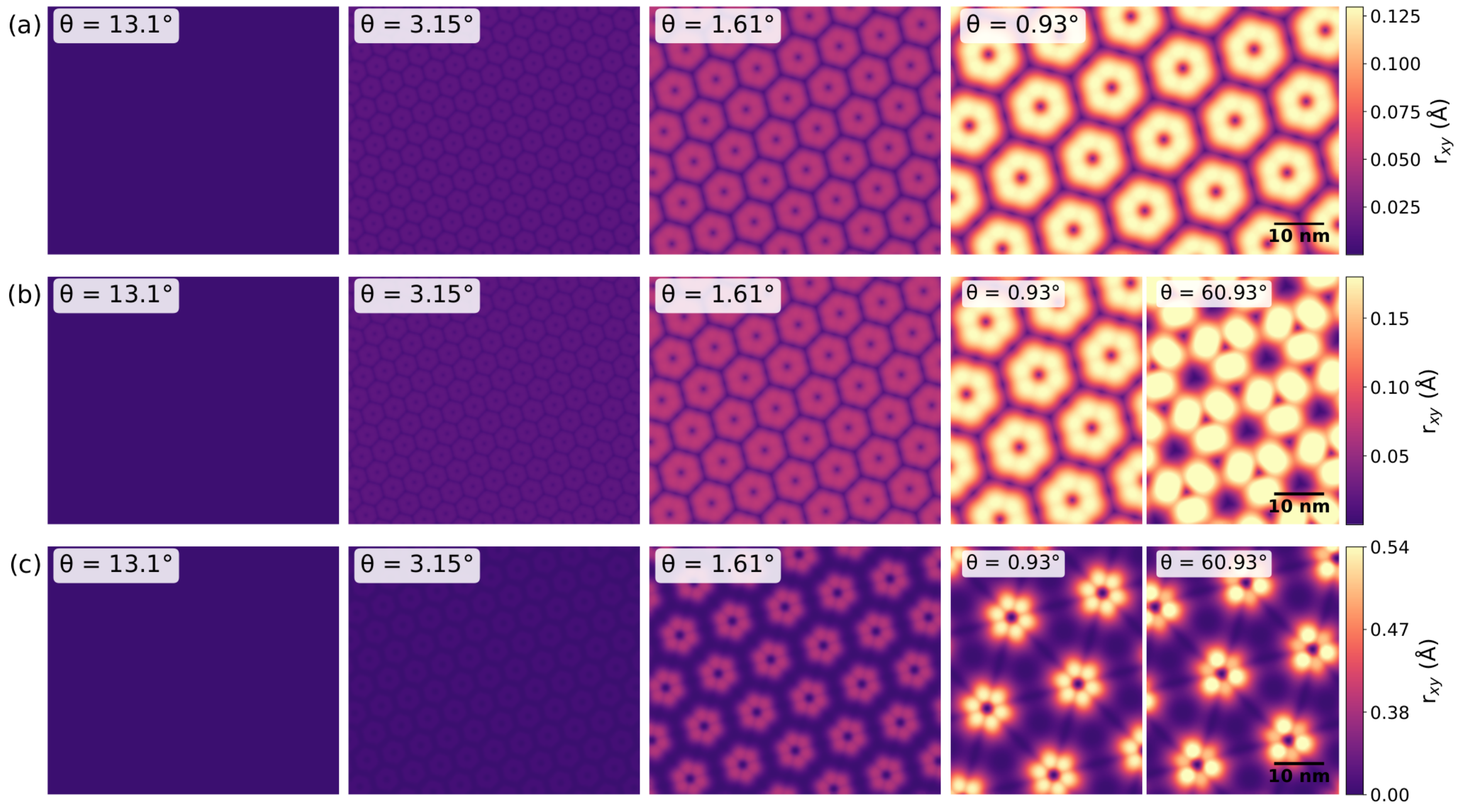}
\caption{In-plane atomic displacement magnitude across the moir\'{e}
unit cell for (a)~graphene, (b)~\textit{h}-BN, and (c)~MoS$_2$ at four
representative twist angles. For 
\textit{h}-BN and MoS$_2$, an additional antiparallel-orientation panel is included as representation.}
\label{fig:intralayer}
\end{figure*}

Figures~\ref{fig:interlayer} and~\ref{fig:intralayer} map the interlayer distance and in-plane atomic displacements, respectively, for all three studied materials at representative twist angles from large to small. The spatial extrema of the interlayer distance, with maxima at AA regions and minima at AB/BA domains (parallel orientation, $-30^\circ < \theta <  30^\circ$) are in close agreement with the equilibrium interlayer separations of the corresponding pristine bilayer stackings reported in Table~\ref{tab:bilayer_distances}, with intermediate regions interpolating smoothly between these values. This broad consistency between moiré-scale relaxation results and benchmark stacking calculations suggests that our fine-tuned models capture the relevant interlayer energetics with reasonable fidelity across the full moiré cell, not only at the pristine systems used for validation. At large twist angles, reconstruction is negligible and both the interlayer distance and in-plane displacements remain nearly uniform, consistent with the rigid-bilayer regime~\cite{yoo2019atomic,weston2020reconstruction}. As $\theta$ decreases, the picture changes dramatically, with the growing moir\'e wavelength making it energetically favorable for the system to expand AB/BA stacking domains at the expense of the higher-energy AA regions. When $\theta$ is small, this process is well advanced, with AB/BA domains occupying the majority of the moir\'e cell and AA domains reduced to small localized spots at the intersections of narrow soliton lines, in qualitative agreement with experimental observations~\cite{yoo2019atomic,weston2020reconstruction}.

How strongly this reconstruction develops depends sensitively on the material. Graphene and \textit{h}-BN behave similarly, as expected from their comparable interlayer energy corrugation (Table~\ref{tab:bilayer_energy}) and nearly equal in-plane stiffness ($E \approx 1~\mathrm{TPa}$~\cite{lee2008measurement} and $\approx 0.865~\mathrm{TPa}$~\cite{falin2017mechanical}, respectively), with \textit{h}-BN showing only a marginally larger reconstruction amplitude. MoS$_2$, however, stands apart, with its substantially reduced in-plane stiffness ($E \approx 270~\mathrm{GPa}$~\cite{bertolazzi2011stretching}) combined with a deeper interlayer energy corrugation (as evidenced by the larger separation between AA and AB equilibrium interlayer distances in Table~\ref{tab:bilayer_distances}) providing both a stronger driving force and a more compliant lattice, producing reconstruction amplitudes significantly larger than those of graphene or \textit{h}-BN at equivalent angles. For \textit{h}-BN and MoS$_2$, the same energy-minimization logic operates in the antiparallel orientation, where the inequivalence of the two sublattice atom types enriches the stacking landscape with new registries (AA$'$, AB$'$, BA$'$) and modifies the domain geometry, though the system again seeks to maximize the area of energetically favorable configurations at the expense of higher-energy regions (Fig.~\ref{fig:interandintralayer_a60}). Note that, in these cases, AA$'$ becomes the most energetically favorable domain instead of AB as in the parallel cases, while AB$'$ has a slightly higher energy and BA$'$ is unfavorable (Table~\ref{tab:bilayer_energy}). In the antiparallel orientation, the inequivalence of the distinct sublattice atom types breaks the symmetry between AB$'$ and BA$'$ registries, which are no longer energetically and structurally equivalent as AB and BA are in the parallel case. Although both \textit{h}-BN and MoS$_2$ exhibit this asymmetry, it manifests considerably more strongly in \textit{h}-BN, as directly visible in Figs.~\ref{fig:interlayer} and~\ref{fig:intralayer} in the rightmost panels. This could be explained by the difference in layer structure between the two materials, which determines how vdW forces are transmitted across the interface. In \textit{h}-BN, the fully planar layer structure allows interlayer vdW interactions to couple the two layers directly, amplifying the registry asymmetry in the reconstructed domain geometry. In MoS$_2$, interlayer vdW interactions are mediated by the outer sulfur atoms rather than the metal sites, reducing the sensitivity of the coupling to the sublattice registry and thereby moderating the AB$'$/BA$'$ asymmetry in the reconstruction. The non-uniform displacement field underlying this reconstruction necessarily implies a spatially heterogeneous distribution of in-plane strain. The AB/BA stacking regions, having relaxed toward ideal registries, are nearly strain-free, while the rapid registry transitions at the soliton lines concentrate large deformation gradients in these narrow regions, and the AA stacking regions accumulate elevated rotational strain at the intersections of the soliton network~\cite{kazmierczak2021strain}. The nature and evolution of this strain landscape are examined quantitatively in the following section.

\subsubsection{Local Strain Distribution}

\begin{figure*}[htbp]
\centering
\includegraphics[width=0.85\textwidth]{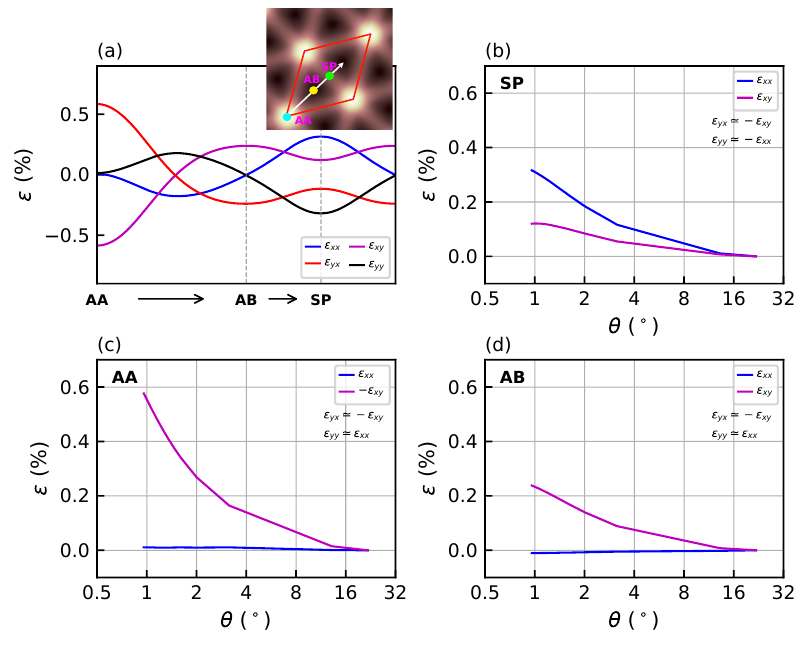}
\caption{Local strain tensor components along the high-symmetry 
path at $\theta = 0.93^\circ$ (a) and twist-angle dependence of 
local strain components in the SP, AA, and AB stacking regions, 
respectively (b--d) for twisted bilayer graphene.}
\label{fig:localstrainsGraphene}
\end{figure*}

\begin{figure*}[htbp]
\centering
\includegraphics[width=0.95\textwidth]{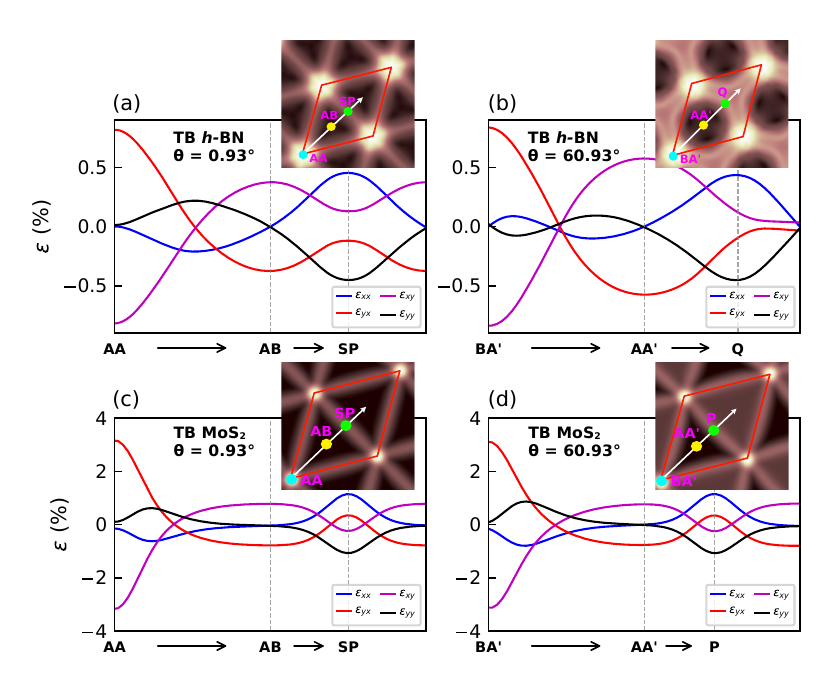}
\caption{Local strain tensor components along the high-symmetry 
path for twisted bilayer \textit{h}-BN in the parallel 
orientation at $\theta = 0.93^\circ$ (a), \textit{h}-BN in the 
antiparallel orientation at $\theta = 60.93^\circ$ (b), 
MoS$_2$ in the parallel orientation at $\theta = 0.93^\circ$ 
(c), and MoS$_2$ in the antiparallel orientation at 
$\theta = 60.93^\circ$ (d).}
\label{fig:localstrainhBNMoS2}
\end{figure*}

The spatially heterogeneous displacement field established by 
atomic reconstruction implies, by definition, a non-uniform 
strain distribution in the moir\'e cell. To characterize 
this distribution quantitatively, we extract the four components 
of the in-plane strain tensor along a representative path 
traversing the distinct stacking environments at $\theta = 
0.93^\circ$ for all three materials 
(Figures~\ref{fig:localstrainsGraphene} 
and~\ref{fig:localstrainhBNMoS2}).

The strain profiles in Fig.~\ref{fig:localstrainsGraphene}(a) reveal distinct features in each stacking region, directly reflecting the underlying energy landscape of the reconstruction. Within the AB stacking domains, the tensor components satisfy $\varepsilon_{yx} \approx -\varepsilon_{xy}$ and $\varepsilon_{xx} \approx \varepsilon_{yy} \approx 0$, a pattern characteristic of nearly pure local rigid rotation with negligible bond distortion. This is precisely what one expects from a region that has successfully relaxed toward the energy minimum of the interlayer potential, where the lattice rotates locally to reduce the interlayer twist toward the commensurate Bernal limit while bond lengths remain close to their equilibrium values. The saddle-point (SP) regions present an entirely different picture, with all four strain components becoming simultaneously significant and spatially varying. These regions mark the boundaries between adjacent AB and BA stacking domains, where the lattice must negotiate a rapid registry transition within a spatially narrow zone. The AB and BA stacking domains on either side of each soliton line rotate in the same direction as they each relax toward their respective energy minima (perfectly Bernal stacking), so that the SP region must accommodate the resulting shear while simultaneously maintaining structural continuity across the boundary, giving rise to the most complex deformation field in the moir\'e cell. The AA stacking regions occupy an intermediate position, retaining a predominantly rotational character but with significantly larger amplitudes than the AB stacking domains. This behavior reflects two compounding effects. The AA stacking corresponds to the energy maximum of the interlayer potential, from which the system seeks to escape but cannot, because the $C_3$ symmetry of the hexagonal lattice topologically requires three soliton lines to meet at a single point~\cite{nguyen2021electronic}, forcing the system to pass through the unfavorable AA registry at each such junction. Furthermore, since the surrounding AB and BA domains all rotate in the same direction as they relax toward their respective energy minima, the AA region is driven to rotate in the opposite direction by geometric continuity, much as a central gear is forced to counter-rotate when all surrounding gears turn in the same sense. The AA stacking regions therefore develop a large local rotation of opposite sign to that of the surrounding domains, yet retain negligible bond distortion. This complete spatial picture, with AB stacking domains remaining nearly undistorted, AA stacking regions developing large local rotations without appreciable bond distortion, and the soliton lines concentrating the most complex deformation, is in direct agreement with the experimental observation of strain field of Kazmierczak \textit{et al.}~\cite{kazmierczak2021strain}, who directly mapped the full strain tensor in reconstructed twisted bilayer graphene using four-dimensional scanning transmission electron microscopy.

The twist-angle dependence (Fig.~\ref{fig:localstrainsGraphene}, panels b--d) reinforces this picture quantitatively. The AB stacking domains remain nearly undistorted over the entire range studied, while the AA stacking regions accumulate progressively larger rotational strain with decreasing angle, as the system is driven further from the unstable AA registry without a nearby stable configuration to relax toward. The SP regions develop the most complex deformation field, in which all four strain components acquire significant magnitudes, reflecting the shear that the lattice must sustain at the soliton lines separating adjacent AB and BA stacking domains. The same spatial pattern extends naturally to \textit{h}-BN and MoS$_2$ in both orientations (Fig.~\ref{fig:localstrainhBNMoS2}). In the parallel orientation, both materials reproduce the same qualitative pattern, with nearly undistorted AB stacking domains, moderate rotational strain at AA stacking areas, and the most complex multi-component deformation at the soliton lines (panels a and c). The strain amplitudes are noticeably larger for MoS$_2$ than for \textit{h}-BN, consistent with the stiffness difference established in the preceding section, with the more compliant MoS$_2$ lattice accommodating the registry transition through larger local distortions.

The antiparallel orientation introduces qualitatively new features for both materials (Fig.~\ref{fig:localstrainhBNMoS2}, panels b and d), yet the same underlying physical picture persists. The additional high-symmetry registries (AA$'$, AB$'$, and BA$'$), with their distinct binding energies (Table~\ref{tab:bilayer_energy}), enrich the stacking landscape and break the left-right symmetry of the strain profiles, as different domains develop distinct strain signatures depending on which registry they relax toward. Unlike the parallel orientation where AB and BA are energetically equivalent, the inequivalence of AB$'$ and BA$'$ means that domains on either side of a transition region relax by different amounts, producing an asymmetric strain landscape that directly reflects the underlying energy asymmetry. Nevertheless, the same underlying logic persists, with the transition regions between stacking domains concentrating the largest deformations while the stacking domains themselves remain comparatively undistorted. Although both \textit{h}-BN and MoS$_2$ possess this asymmetry, it is considerably more pronounced in \textit{h}-BN, as directly visible in panels b and d of Fig.~\ref{fig:localstrainhBNMoS2}. This difference could be explained by the same screening effect discussed in the preceding section, where the fully planar layer structure of \textit{h}-BN may transmit the registry asymmetry more directly across the interface than in MoS$_2$, where interlayer vdW interactions are mediated by the outer sulfur atoms, potentially reducing the sensitivity of the coupling to the sublattice registry and moderating the observed asymmetry. For MoS$_2$, the strain amplitudes at the transition regions reach several percent, substantially larger than in \textit{h}-BN at equivalent angles, consistent with the greater mechanical compliance and deeper interlayer energy corrugation of the MoS$_2$ lattice. Detailed strain profiles for \textit{h}-BN and MoS$_2$ in the parallel and antiparallel orientations are provided in Figs.~\ref{fig:localstrain_hbn_a0}, \ref{fig:localstrain_mos2_a0}, \ref{fig:localstrain_hbn_a60}, and~\ref{fig:localstrain_mos2_a60} of the Supplementary Materials.

\begin{figure*}[htbp]
\centering
\includegraphics[width=1.0\textwidth]{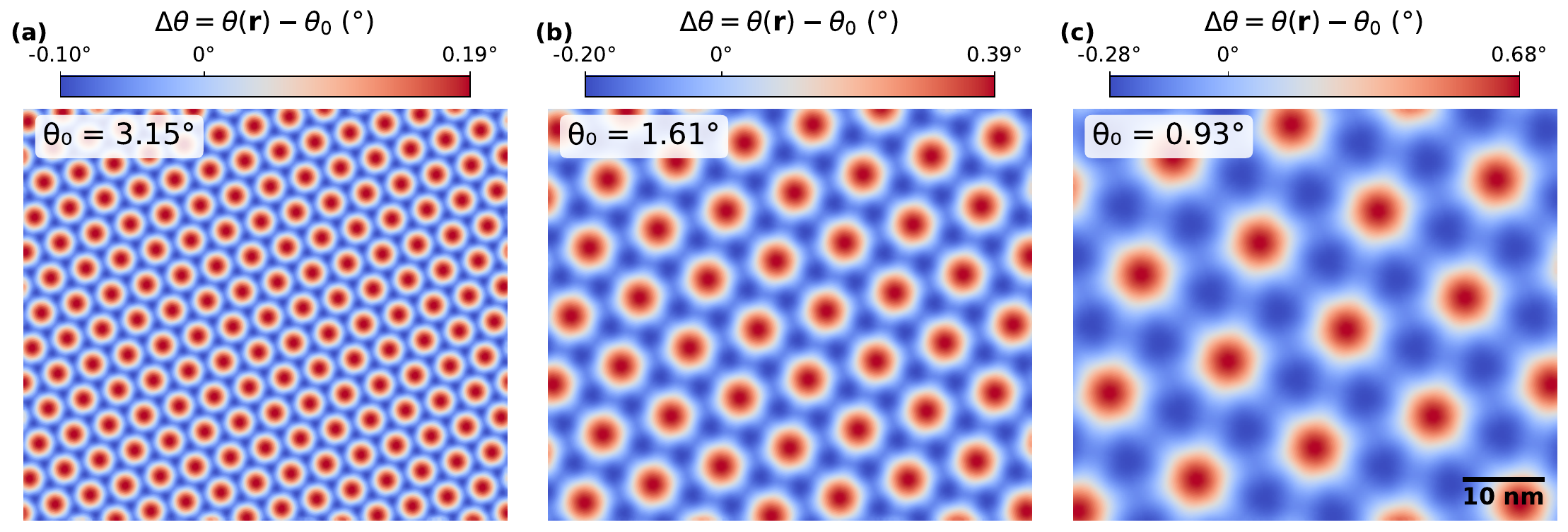}
\caption{Spatial maps of the local twist angle deviation $\Delta\theta(\mathbf{r}) = \theta(\mathbf{r}) - \theta_0$ for twisted bilayer graphene at (a)~$\theta_0 = 3.15^\circ$, (b)~$\theta_0 = 1.61^\circ$, and (c)~$\theta_0 = 0.93^\circ$. Negative values (blue) indicate that reconstruction has reduced the local rotation below the nominal twist angle, as in the AB domains; positive values (red) reflect an increase above $\theta_0$, concentrated at the AA stacking zones. The growing colorbar range with decreasing angle reflects the progressive intensification of reconstruction.}
\label{fig:localtwistangle}
\end{figure*}

This picture is further substantiated by the local twist angle maps shown in Fig.~\ref{fig:localtwistangle}, which display the spatially resolved deviation $\Delta\theta(\mathbf{r}) = \theta(\mathbf{r}) - \theta_0$ of the local interlayer rotation from the nominal twist angle, computed from the nearest-neighbor bond vectors of equivalent sublattice atoms between the two layers. Note that atomic reconstruction takes place as the most energetically favorable stacking domains are maximized and accordingly, the other domains undergo local deformation. The final optimization is obtained by the energy-minimizing compromise between these two processes. The maps reveal a striking spatial pattern: the AB stacking domains exhibit $\Delta\theta < 0$, as reconstruction drives the local rotation toward the commensurate Bernal limit, while the AA stacking zones develop $\Delta\theta > 0$, rotating in the opposite sense as a geometric consequence of the surrounding domain network~\cite{kazmierczak2021strain}. The former is the direct consequence of the discussed maximization of energetically favorable domains. This spatially structured variation of the local twist angle is an intrinsic consequence of atomic reconstruction and should not be confused with twist-angle disorder~\cite{beechem2014rotational, wilson2020disorder}, the extrinsic spatial inhomogeneity of the global twist angle that arises from imprecision in the fabrication process and varies on micrometre length scales. The former is a deterministic, periodically structured response to the moir\'e energy landscape; the latter is a stochastic deviation from the intended global twist angle. As $\theta$ decreases and reconstruction deepens, both effects intensify, with the AB stacking domains approaching ever closer to zero local twist and the AA stacking areas developing increasingly large positive deviations. Notably, experimental measurements have shown that the AA rotation does not grow indefinitely but saturates at a finite value as $\theta$ approaches zero, bounded by the surrounding soliton network~\cite{kazmierczak2021strain}, a trend consistent with the behavior observed in Fig.~\ref{fig:localtwistangle}.

Extending the analysis to lower twist angles, where the moir\'e supercells become computationally demanding even for fine-tuned MACE due to memory constraints, we use the REBO+KC classical force field as a qualitative guide, justified by the reasonable agreement it shows with fine-tuned MACE at accessible angles (Fig.~\ref{fig:strainAAABMACEnREBO}). It should be noted, however, that this qualitative agreement in reconstruction trends does not imply that REBO+KC is a reliable tool for quantitative predictions. As demonstrated in Section~\ref{sec:phonon}, it systematically overestimates strain amplitudes by a factor of several and yields phonon spectra in poor agreement with experiment, precisely the deficiencies that motivate the use of fine-tuned MACE in the present work. Within this caveat, the REBO+KC calculations at $\theta \approx 0.2^\circ$ (Fig.~\ref{fig:localtwistangle_REBOKC}) suggest that the local twist angle in AB domains can reach $0^\circ$ (i.e., prefect Bernal stacking) with decreasing angle, the local deformation in other stacking domains is still significant.

%while the AA stacking region rotation saturates at a finite value bounded by the surrounding soliton network, again consistent with the experimental observations of Kazmierczak \textit{et al.}~\cite{kazmierczak2021strain} and further supporting the physical picture established above. 

\begin{figure*}[htbp]
\centering
\includegraphics[width=0.65\textwidth]{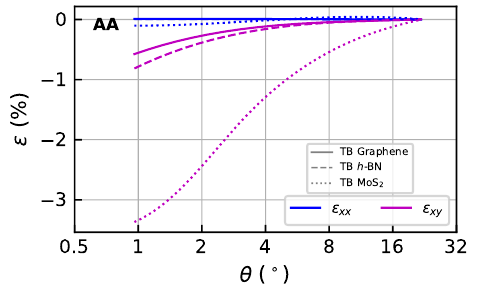}
\caption{Twist-angle dependence of the strain components in the AA stacking region for twisted bilayer graphene, \textit{h}-BN, and MoS$_2$.}
\label{fig:strainAAallmaterials}
\end{figure*}

The material dependence of the strain landscape is summarized in Fig.~\ref{fig:strainAAallmaterials}, which compares the twist-angle evolution of the AA-region strain for all three materials. The ordering is fully consistent with the relative mechanical compliance of the three materials discussed in Section~\ref{sec:reconstruction}, with MoS$_2$ developing strain amplitudes significantly larger than graphene or \textit{h}-BN at equivalent angles, while graphene and \textit{h}-BN track each other closely, with \textit{h}-BN showing a marginal excess.

\begin{figure*}[t]
\centering
\includegraphics[width=1.0\textwidth]{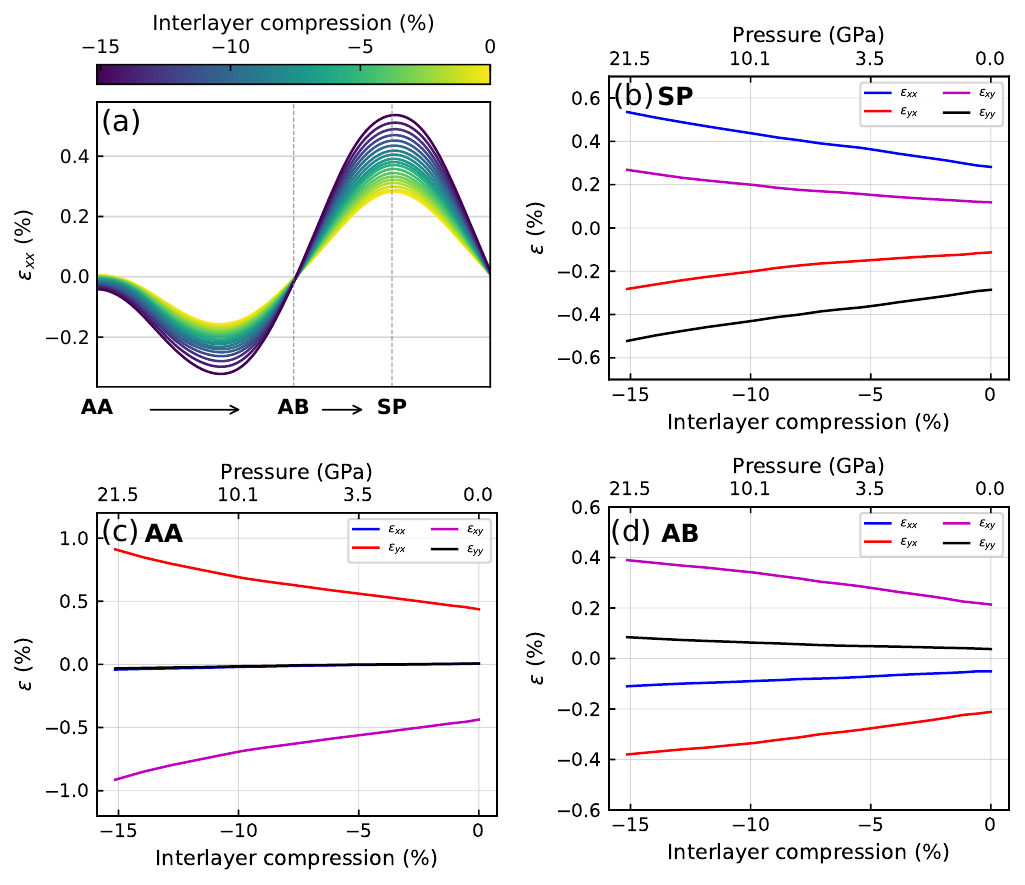}
\caption{Local strain tensor components in twisted bilayer graphene at $\theta = 2^\circ$ under increasing out-of-plane compression, for (a)~the spatial profile along the AA$\to$AB$\to$SP path and (b--d)~strain tensor components as a function of interlayer compression (applied pressure) in different stacking regions in the moiré cell.}
\label{fig:strainunderpressure}
\end{figure*}

Vertical pressure is known as a practical means of tuning interlayer coupling in van der Waals bilayers, thereby offering a route to flat-band formation at large twist angles ~\cite{yankowitz2019tuning,carr2018pressure}. The atomic reconstruction discussed above can furthermore be enlarged by such external mechanical pertubation. Here, we model vertical compression using ideal power-law potential walls added to the MACE energy, which creates the opposite forces (in the vertical direction) on top and bottom layers. As compression increases, the spatial strain profile evolves systematically (Fig.~\ref{fig:strainunderpressure}a), with the AA and AB stacking domains develop stronger local rotations while retaining negligible bond distortion, and the strain amplitude at the SP regions grows by nearly a factor of two over the full compression range studied, directly demonstrating the selective concentration of pressure-induced deformation at the registry-transition zones. As compression reduces the interlayer separation, the vdW interaction between the layers strengthens, increasing the energy differences between stacking registries that differ in their local atomic arrangement. The driving force for reconstruction therefore intensifies, and since the SP regions are where registry transitions are most rapid and the energy gradient is largest, they bear a disproportionate share of the additional deformation, while the AB stacking domains, already relaxed near their energy minimum, remain comparatively undisturbed. The strain field is therefore a continuously tunable landscape whose amplitude reflects the full interplay between interlayer energetics, in-plane elasticity, and external mechanical conditions. Note that the  increase in interlayer electronic coupling (as interlayer distance decreases) and the enhancement of atomic reconstruction together account for flat-band formation and superconducting at larger twist angles ~\cite{yankowitz2019tuning}.

\subsection{Moiré Effects on Lattice Dynamics}
\label{sec:phonon}
\begin{figure*}[htbp]
\centering
\includegraphics[width=\textwidth]{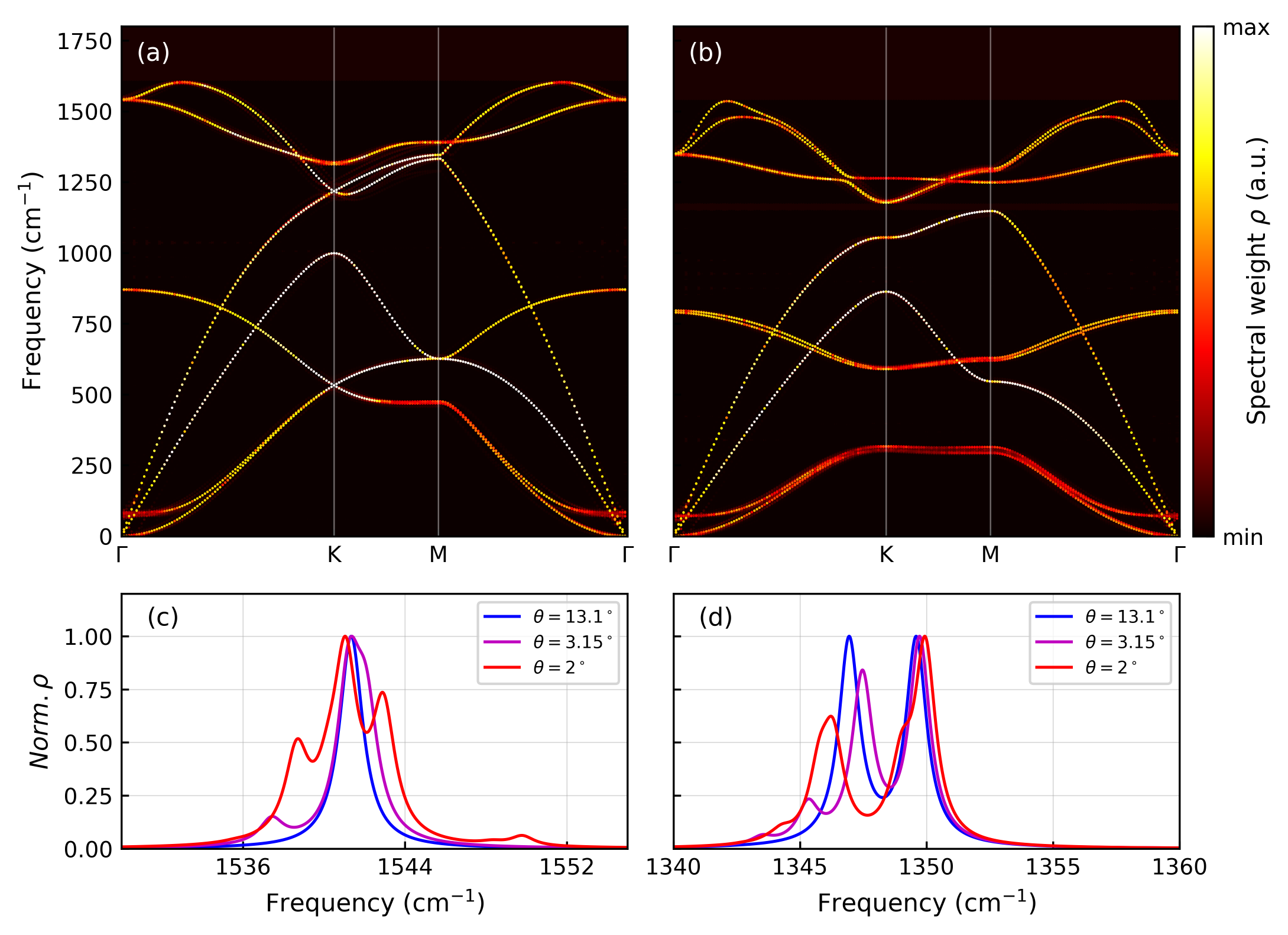}
\caption{Unfolded phonon dispersion and G-band spectral weight for twisted bilayer graphene and \textit{h}-BN. (a,b)~Full phonon dispersion weighted by spectral weight $\rho$ at $\theta = 2^\circ$ for graphene and \textit{h}-BN, respectively. (c,d)~Normalized spectral weight at $\Gamma$ in the G-band region for three representative twist angles.}
\label{fig:grapheneandhBN_phonon}
\end{figure*}

The spatially structured strain field established by atomic reconstruction characterized in the preceding section couples directly to the lattice dynamics of the moir\'e superlattice. Acting through the Gr\"uneisen mechanism~\cite{mohiuddin2009uniaxial}, local in-plane strain breaks the rotational symmetry of the hexagonal lattice, lifting mode degeneracies and redistributing spectral weight across the phonon spectrum in a manner that encodes the underlying domain structure. To isolate these reconstruction-induced signatures from the dense band-folding background of the moir\'e supercell, we analyze the phonon response through the unfolded spectral weight defined in Eq.~(\ref{eq:spectral_weight}) (as illustrated in Fig.S7), which projects each supercell eigenmode onto primitive-cell Bloch character and thereby recovers the twist-angle evolution of individual phonon branches, providing a basis for comparison with experimental Raman spectra \cite{gadelha2021localization}. 

\begin{figure*}[!htbp]
\centering
\includegraphics[width=0.95\textwidth]{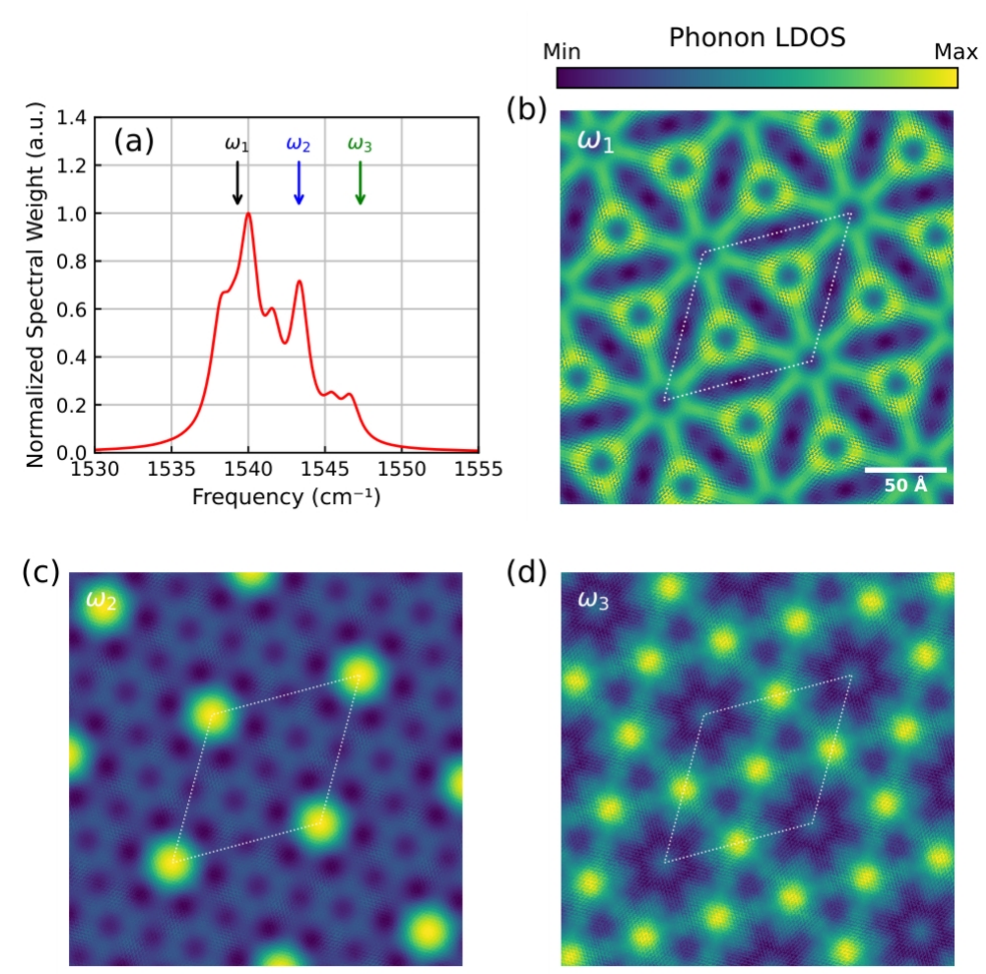}
\caption{Local phonon density of states (LPDOS) in the G-band region 
for twisted bilayer graphene at $\theta = 1.4^\circ$. 
(a)~Total normalized spectral weight as a function of 
frequency, with three marked frequencies $\omega_1$, 
$\omega_2$, and $\omega_3$. (b--d)~Real-space maps of LPDOS at frequencies $\omega_1$, $\omega_2$, and $\omega_3$, 
respectively and at the $\Gamma$ point, with the moiré unit cell indicated by the white 
dashed outline.}
\label{fig:localphonon}
\end{figure*}

Figures~\ref{fig:grapheneandhBN_phonon}(a,b) show the full unfolded phonon dispersions at $\theta = 2^\circ$ for twisted bilayer graphene and \textit{h}-BN, respectively. The reconstruction signature concentrates at the zone center, where the local strain field generated across the moiré cell breaks the $D_{3h}$ symmetry of the hexagonal lattice and lifts the $E_{2g}$ degeneracy, in direct analogy with what uniaxial strain produces in strained monolayer and bilayer systems~\cite{mohiuddin2009uniaxial, huang2009phonon, androulidakis2018strained, conley2013bandgap, sohier2017breakdown}. Figures~\ref{fig:grapheneandhBN_phonon}(c,d) track this evolution through the spectral weight at $\Gamma$: in both materials, the in-plane optical spectral weight progressively broadens and develops an increasingly asymmetric profile with decreasing twist angle. The spectral weight is a mode-projected quantity that measures how much each supercell eigenmode retains primitive-cell Bloch character. Prior experimental work~\cite{gadelha2021localization} established that the G-band splitting in reconstructed twisted bilayer graphene is only accessible to nano-Raman, whose probe spot is smaller than the soliton width, while micro-Raman averages over the full moiré cell and cannot resolve it. The micro-Raman G-band linewidth reaches at most $\sim\!19$~cm$^{-1}$ near the magic angle~\cite{gadelha2021localization}, with no resolved splitting. More recent micro-Raman measurements~\cite{solanki2025anomalies} do detect a small splitting near $\theta \sim 1^\circ$, but its magnitude barely exceeds $\sim\!10$~cm$^{-1}$. This experimental picture stands in contrast with the $\sim\!70 - 75$~cm$^{-1}$ splitting predicted at $\theta \approx 1^\circ$ by classical force-field calculations using the REBO+KC potential~\cite{lamparski2020soliton,gadelha2021localization}, a value large enough that micro-Raman should have resolved it unambiguously, yet it has not been observed. A natural explanation emerges from the strain analysis in Fig.~\ref{fig:strainAAABMACEnREBO}, where the strain amplitude predicted by the REBO+KC potential is three to four times larger than that obtained with our fine-tuned MACE, and since G-band splitting scales with local strain~\cite{mohiuddin2009uniaxial}, this difference propagates directly into the phonon spectrum. The fine-tuned MACE potential, by faithfully reproducing the DFT strain landscape, yields splitting amplitudes of $\sim\!8$~cm$^{-1}$ at $\theta = 2^\circ$ (Fig.~\ref{fig:grapheneandhBN_phonon}c) and $\sim\!10$~cm$^{-1}$ at $\theta = 1.4^\circ$ (Fig.~\ref{fig:localphonon}a), consistent with the experimental bounds. The analogous spectral weight evolution in twisted bilayer \textit{h}-BN (Fig.~\ref{fig:grapheneandhBN_phonon}d) follows the same physical logic, but a direct experimental comparison is currently not possible owing to the nonresonant nature of Raman scattering in \textit{h}-BN, which yields signals too weak for reliable nano-Raman measurements, and high-resolution data on reconstructed twisted \textit{h}-BN bilayers have not yet been reported. The spectral weight evolution presented here therefore constitutes a theoretical prediction awaiting experimental verification.

The phonon signatures discussed above reflect contributions 
from the entire moiré cell; the underlying strain field, 
however, is spatially structured, and so too is the vibrational 
response. Figure~\ref{fig:localphonon} probes this spatial texture directly for twisted bilayer graphene at $\theta = 1.4^\circ$, resolving three discernible features $\omega_1 < \omega_2 < \omega_3$ in the G-band spectral weight and mapping their real-space phonon local density of states. At $\omega_1$ the phonon LDOS is localized predominantly at the AB stacking domains with a weaker contribution at the AA stacking areas, and at $\omega_3$ it concentrates along the soliton lines consistent with the nano-Raman hyperspectral imaging of Gadelha~\textit{et al.}~\cite{gadelha2021localization}, who identified analogous spatial differentiation between the lower- and higher-frequency G satellites in reconstructed twisted bilayer graphene. Our calculations additionally resolve an intermediate feature at $\omega_2$ that is absent in that work, whose LDOS map concentrates at the AA stacking zones. A direct quantitative comparison is nonetheless not warranted: our results are obtained at $\theta = 1.4^\circ$, whereas those measurements were performed at $\theta \approx 0.09^\circ$, where reconstruction is far stronger and spatial contrasts far sharper. However, the qualitative correspondence in both the spectral and spatial 
dimensions provides independent support for the physical 
picture delivered by the fine-tuned MACE potential.

\begin{figure*}[!htbp]
\centering
\includegraphics[width=\textwidth]{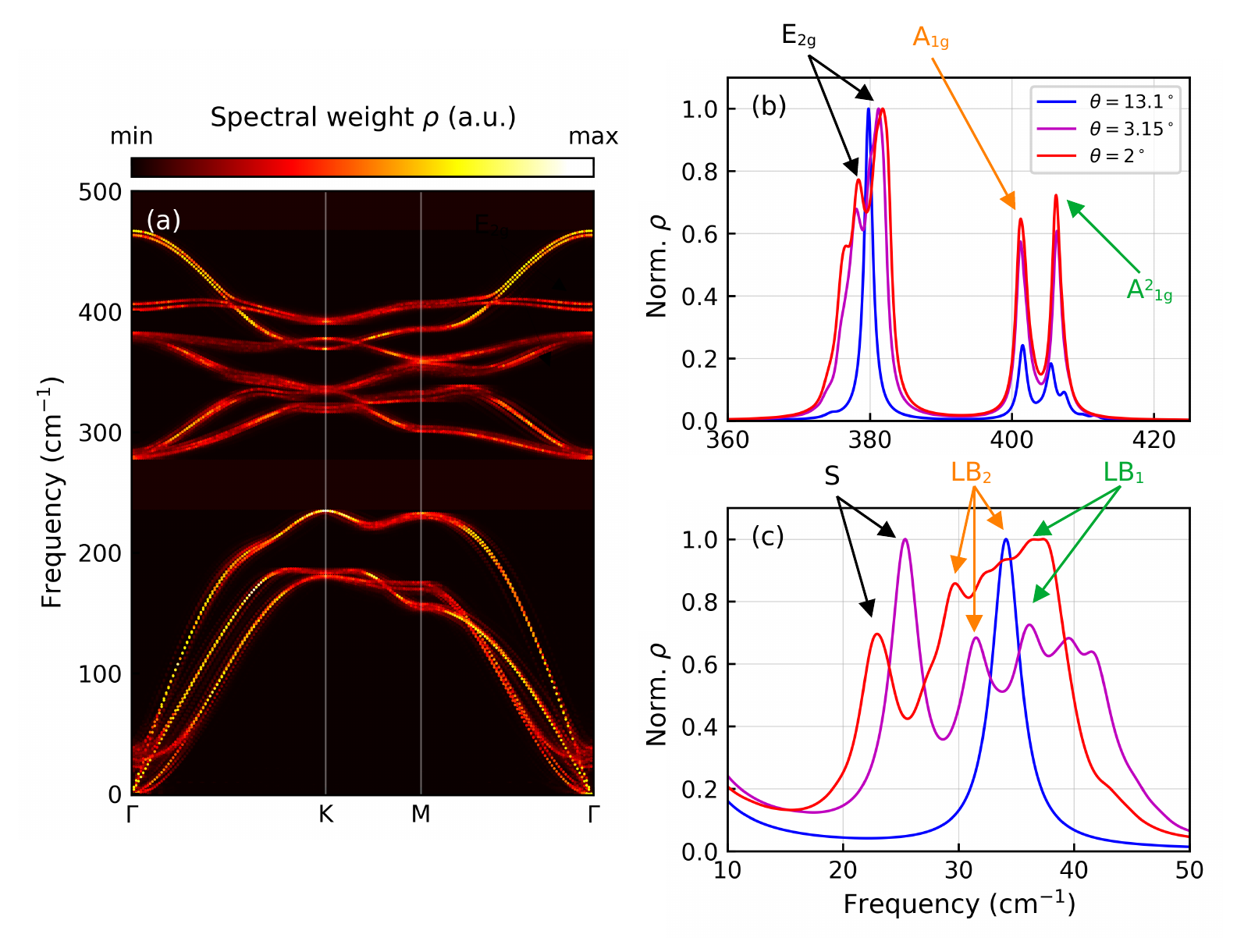}
\caption{Unfolded phonon spectral weight for twisted bilayer MoS$_2$. (a)~Full phonon dispersion weighted by spectral weight $\rho$ at $\theta = 2^\circ$. (b)~Normalized high-frequency spectral weight in the E$_{2g}$/A$_{1g}$ region for three representative twist angles. (c)~Normalized low-frequency spectral weight in the interlayer shear (S) and layer-breathing (LB) region at the same angles.}
\label{fig:mos2_phonon}
\end{figure*}

The vibrational response of twisted bilayer MoS$_2$ follows 
the same reconstruction-driven logic. Figure~\ref{fig:mos2_phonon}(a) shows its full unfolded 
phonon dispersion, with 
panels~(b) and~(c) resolving the spectral weight evolution 
in the high- and the low-frequency 
region, respectively, for three representative twist angles.

Among the high-frequency modes, the A$_{1g}$ mode near 402~cm$^{-1}$, in which the two sulfur atoms within each unit cell vibrate out of plane in opposite directions, remains essentially unshifted over all angles studied. As a purely out-of-plane motion, it couples only weakly to the in-plane strain field generated by reconstruction and therefore serves as a natural internal frequency reference. It is worth noting that a second out-of-plane mode of A$_{1g}^2$ symmetry, visible at higher frequency in our phonon dispersion, is Raman-inactive and therefore not observable in Raman spectroscopy~\cite{lee2010anomalous}. The E$_{2g}$ mode near 381~cm$^{-1}$, involving in-plane relative motion of the Mo and S atoms, is by contrast the most sensitive to local strain: as reconstruction deepens with decreasing angle, its spectral weight broadens and develops an increasingly asymmetric profile (Fig.~\ref{fig:mos2_phonon}b), reflecting the same symmetry-breaking mechanism established above for graphene and \textit{h}-BN. The low-frequency interlayer modes provide a complementary perspective (Fig.~\ref{fig:mos2_phonon}c). At $\theta = 13.1^\circ$, the shear (S) mode, in which the two layers slide against each other in the plane ($\sim\!24$~cm$^{-1}$), is absent, while the layer-breathing (LB) mode, corresponding to out-of-plane expansion and compression between the layers ($\sim\!33$~cm$^{-1}$), appears as a single well-defined feature. As the twist angle decreases and reconstruction develops, the S mode emerges and strengthens, while the LB mode splits into two distinct branches, a lower branch LB$_2$ ($\sim\!31$~cm$^{-1}$) and an upper branch LB$_1$ ($\sim\!37$~cm$^{-1}$), whose separation grows progressively as reconstruction deepens. These features are in good agreement with the experimental observations of Quan \textit{et al.}~\cite{quan2021phonon}, providing further validation of the fine-tuned MACE potential throughout the full phonon spectrum of reconstructed MoS$_2$ twisted bilayers.

\section{Conclusions} 

In this work, we demonstrate that fine-tuning a foundation MACE model is essential for the predictive simulation of moir\'e systems. While broad-scale foundation models excel at chemical transferability, they systematically fail to resolve the subtle stacking energy differences and interlayer coupling that govern reconstruction, often placing materials in qualitatively incorrect bonding regimes. By fine-tuning on compact datasets built specifically for twisted bilayer materials, incorporating the relevant stacking registries, interlayer spacings, and thermally distorted configurations, we achieve near-DFT accuracy in energies, forces, relaxed structures, and phonon spectra consistently for all three materials studied. This precision enables us to access the low-angle regime, where atomic reconstruction is most pronounced and DFT becomes computationally prohibitively expensive. Our models reveal a consistent reconstruction-induced strain landscape, in which energetically favorable stacking regions relax toward ideal registries, while unfavorable high-symmetry regions, topologically constrained within the soliton network, undergo large local rotations. The soliton lines act as sites of intense deformation gradients, driven by the intrinsic energy landscape rather than extrinsic fabrication disorder. Furthermore, we show that this strain landscape is continuously tunable: vertical compression selectively amplifies strain at the registry-transition regions, increasing the versatility of fine-tuned MLIPs beyond simple twist-angle variations alone. Critically, the predicted strain landscape—characterized by symmetry breaking, spectral weight redistribution, and real-space phonon localization—aligns with experimental Raman observations, providing a level of fidelity that classical force-field potentials, which overestimate local strain by orders of magnitude, cannot achieve. Critically, the predicted strain landscape and phonon signatures align with experimental Raman observations, providing a level of fidelity that classical force-field potentials, which overestimate local strain by orders of magnitude, cannot achieve. While current constraints regarding GPU memory limit simulations at extremely small twist angles, rapid advancements in both hardware capabilities and MLIP architectures are likely to overcome this bottleneck. Moving forward, integrating these models with machine-learning Hamiltonians [64, 65] promises an end-to-end workflow for the fully self-consistent characterization of the structural, vibrational, and electronic properties of moir\'e systems from a single fine-tuned model. Ultimately, the methodology developed here provides a scalable blueprint for exploring the complex energetics of any twisted bilayer or van der Waals heterostructure.

% Acknowledgments
\section*{Acknowledgments} 
The authors acknowledge financial support from the Fédération Wallonie-Bruxelles through the ARC grant ``DREAMS'' (No. 21/26-116), from the EOS project ``CONNECT'' (No. 40007563), from the EU Pathfinder project ``FLATS'' (No.101099139), and from the Belgium FRS-FNRS through the research project (No. T.029.22F). G.-M.R. is Research Director of the Fonds de la Recherche Scientifique - FNRS. Computational resources have been provided by the supercomputing facilities of the Université catholique de Louvain (CISM/UCL) and the Consortium des Équipements de Calcul Intensif en Fédération Wallonie Bruxelles (CÉCI) funded by the Fond de la Recherche Scientifique de Belgique (F.R.S.-FNRS) under convention No. 2.5020.11 and by the Walloon Region. The present research also benefited from computational resources made available on Lucia, the Tier-1 supercomputer of the Walloon Region, infrastructure funded by the Walloon Region under the Grant Agreement No. 1910247.

% % Bibliography
\bibliographystyle{unsrt}
\bibliography{Bibliography}

% Supplementary materials
\clearpage
\newpage
\renewcommand{\thefigure}{S\arabic{figure}}
\setcounter{page}{1}    
\setcounter{figure}{0}    
\renewcommand{\thetable}{S\arabic{table}}
\setcounter{table}{0}

\noindent{\Large \textbf{\underline{Supplementary Materials:}}}\\
\\
\\
\noindent\textbf{A. Model Validation}\\
\\
\noindent\textit{A.1. Pristine Systems}
\captionof{table}{Relative stacking energies $\Delta E$ (meV/atom) for pristine bilayer graphene, \textit{h}-BN, and MoS$_2$ at high-symmetry stacking configurations, compared between PBE-D3 (reference), foundation MACE (OMAT-D3), and fine-tuned MACE (ft-MACE). For each method, $\Delta E$ is computed with respect to the most stable stacking configuration of PBE-D3. Values in parentheses indicate the absolute deviation from the PBE-D3 total energy at the same configuration. Antiparallel configurations (AA$'$, AB$'$, BA$'$), highlighted in red, are specific to \textit{h}-BN and MoS$_2$.}
\label{tab:bilayer_energy}
\begin{ruledtabular}
\begin{tabular}{llccccc}
Material & Method & AA & AB &
\textcolor{red}{AA$'$} &
\textcolor{red}{AB$'$} &
\textcolor{red}{BA$'$} \\
\hline
\multirow{3}{*}{Graphene}
 & PBE-D3  & $16$ & $0$ & --- & --- & --- \\
 & OMAT-D3 & $141$ {\footnotesize$(125)$} & 
   $146$ {\footnotesize$(146)$} & --- & --- & --- \\
 & ft-MACE & $15$ {\footnotesize$(1)$} & 
   $0$ {\footnotesize$(0)$} & --- & --- & --- \\
\hline
\multirow{3}{*}{\textit{h}-BN}
 & PBE-D3  & $19$ & $0$ & $0$ & $4$ & $17$ \\
 & OMAT-D3 & $266$ {\footnotesize$(247)$} & 
   $266$ {\footnotesize$(266)$} & 
   $262$ {\footnotesize$(262)$} & 
   $269$ {\footnotesize$(265)$} & 
   $267$ {\footnotesize$(250)$} \\
 & ft-MACE & $29$ {\footnotesize$(10)$} & 
   $13$ {\footnotesize$(13)$} & 
   $9$ {\footnotesize$(9)$} & 
   $16$ {\footnotesize$(12)$} & 
   $29$ {\footnotesize$(12)$} \\
\hline
\multirow{3}{*}{MoS$_2$}
 & PBE-D3  & $91$ & $0$ & $1$ & $19$ & $87$ \\
 & OMAT-D3 & $377$ {\footnotesize$(286)$} & 
   $333$ {\footnotesize$(333)$} & 
   $330$ {\footnotesize$(329)$} & 
   $332$ {\footnotesize$(313)$} & 
   $377$ {\footnotesize$(290)$} \\
 & ft-MACE & $90$ {\footnotesize$(1)$} & 
   $4$ {\footnotesize$(4)$} & 
   $2$ {\footnotesize$(1)$} & 
   $5$ {\footnotesize$(14)$} & 
   $89$ {\footnotesize$(2)$} \\
\end{tabular}
\end{ruledtabular}

\vspace{12cm}
\captionof{table}{Registry coverage from the stacking order parameter for each twisted bilayer material system, computed over all frames in the training, test, and combined datasets. Percentages are atom-weighted, and $\langle R \rangle$ denotes the mean stacking order parameter value.}
\label{tab:registry_coverage}
\begin{ruledtabular}
\begin{tabular}{llccccccc}
System & Split & Frames & Atoms & AA-like & AB/BA-like & 
SP-like & Intermediate & \(\langle R\rangle\) \\
\hline
\multirow{3}{*}{Graphene}
 & Train & 230 & 81780 & 6.5 & 13.2 & 47.7 & 32.5 & 0.324 \\
 & Test  & 26  & 8172  & 8.3 & 14.2 & 45.8 & 31.6 & 0.334 \\
 & All   & 256 & 89952 & 6.7 & 13.3 & 47.6 & 32.5 & 0.325 \\
\hline
\multirow{3}{*}{\textit{h}-BN}
 & Train & 136 & 32968 & 6.7 & 11.7 & 51.7 & 29.8 & 0.322 \\
 & Test  & 16  & 4312  & 7.1 & 12.6 & 53.6 & 26.7 & 0.314 \\
 & All   & 152 & 37280 & 6.8 & 11.8 & 52.0 & 29.4 & 0.322 \\
\hline
\multirow{3}{*}{MoS$_2$}
 & Train & 160 & 48228 & 4.5 & 11.9 & 44.2 & 39.4 & 0.333 \\
 & Test  & 18  & 4608  & 9.9 & 9.5  & 48.7 & 31.9 & 0.361 \\
 & All   & 178 & 52836 & 5.0 & 11.7 & 44.6 & 38.7 & 0.336 \\
\end{tabular}
\end{ruledtabular}

\bigskip
\bigskip

\begin{center}
\includegraphics[width=0.9\textwidth]
{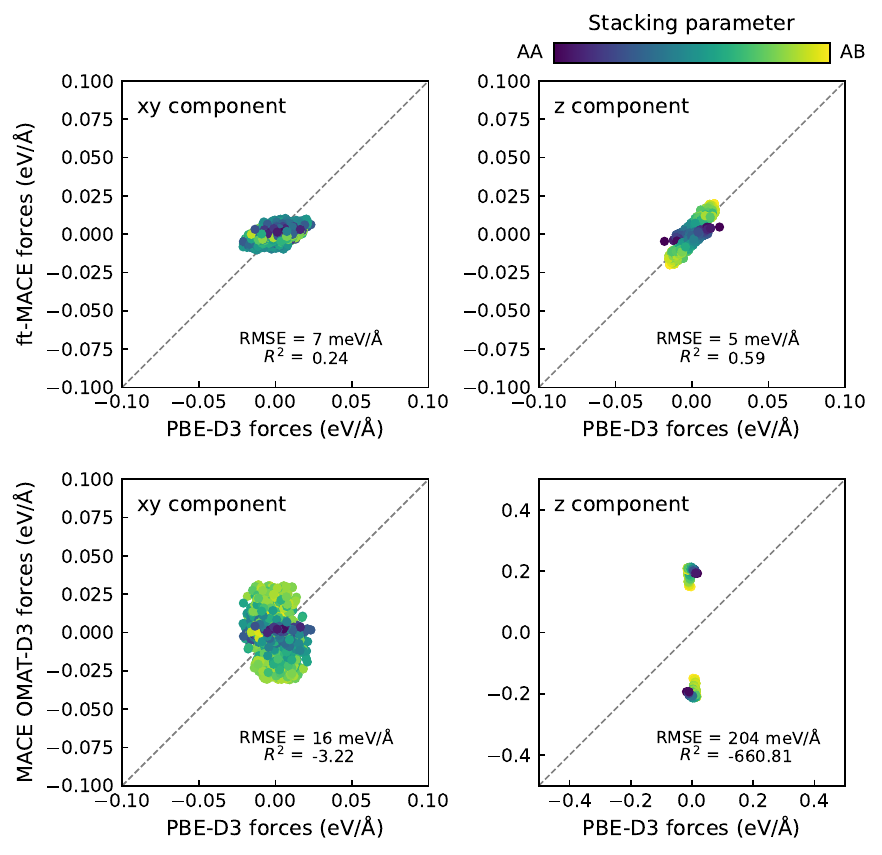}
\end{center}
\captionof{figure}{Parity plots of foundation MACE (OMAT-D3, bottom row) and fine-tuned MACE (ft-MACE, top row) forces against PBE-D3 reference values for twisted bilayer graphene at $\theta \approx 3.1^\circ$, resolved into in-plane ($xy$) and out-of-plane ($z$) components. Data points are colored by the local stacking order parameter, ranging from AA-like (purple) to AB-like (yellow).} \label{fig:stackingparams_tblg}

\bigskip
\bigskip

\begin{center}
\includegraphics[width=0.9\textwidth]
{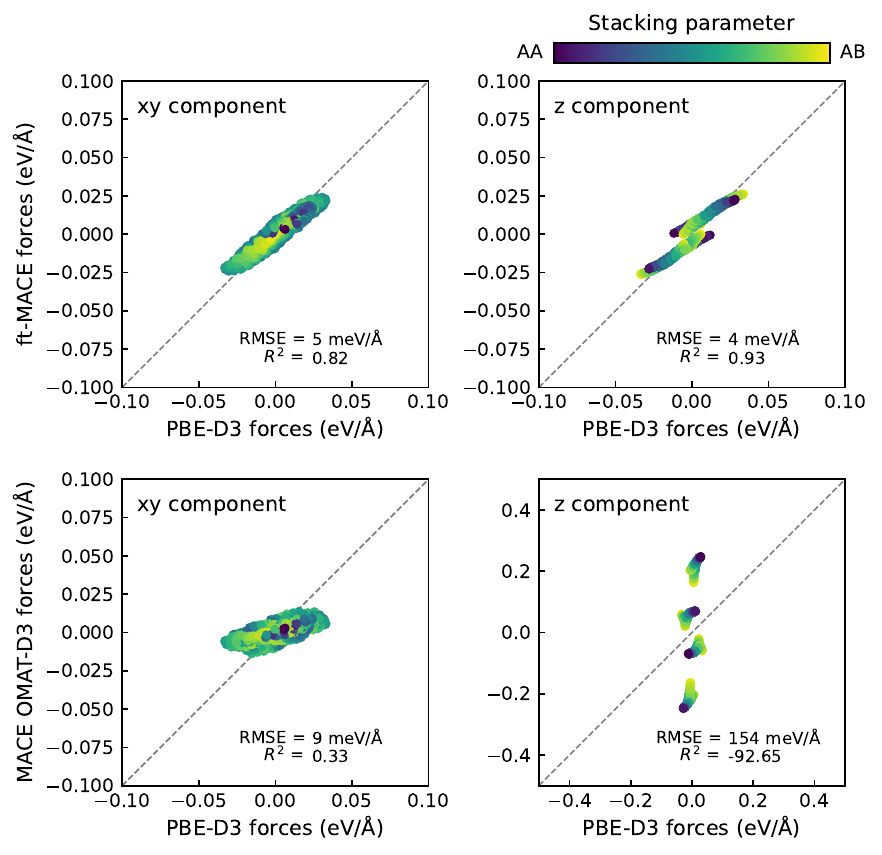}
\end{center}
\captionof{figure}{Same as Fig.~\ref{fig:stackingparams_tblg} for twisted bilayer \textit{h}-BN at $\theta \approx 3.1^\circ$.}
\label{fig:stackingparams_hbn}

\bigskip
\bigskip

\begin{center}
\includegraphics[width=0.9\textwidth]
{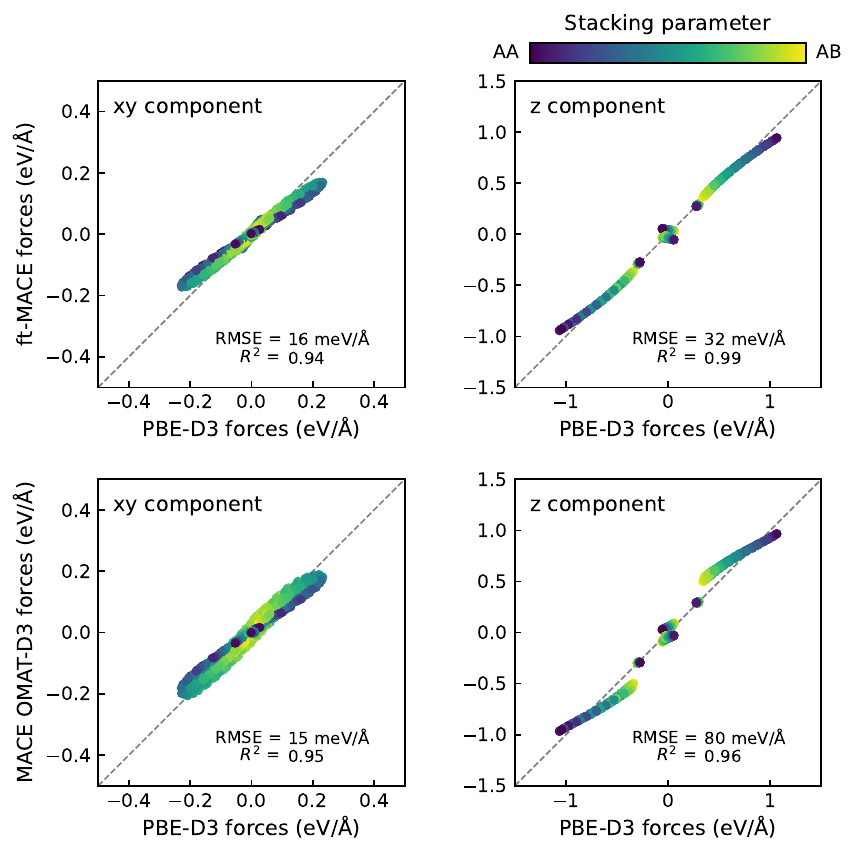}
\end{center}
\captionof{figure}{Same as Fig.~\ref{fig:stackingparams_tblg} for twisted bilayer MoS$_2$ at $\theta \approx 3.1^\circ$.}
\label{fig:stackingparams_mos2}

\newpage

\noindent\textit{A.2. Moir\'e Systems}\\

\begin{center}
\includegraphics[width=0.9\textwidth]
{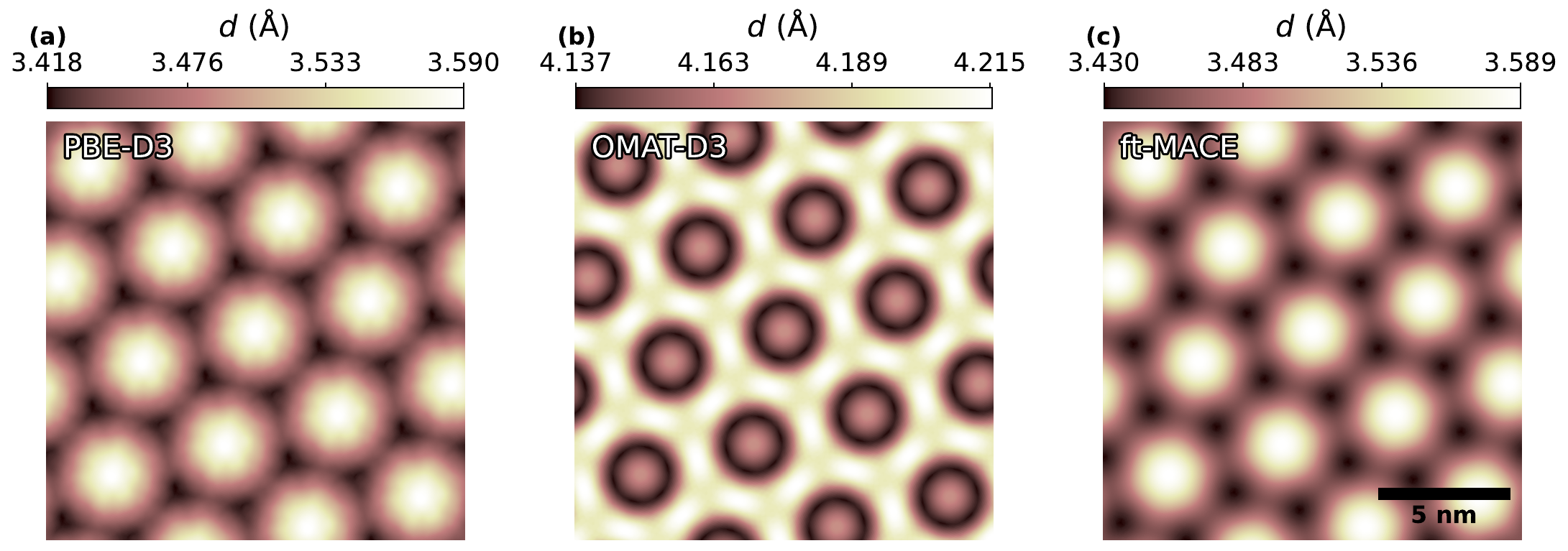}
\end{center}
\captionof{figure}{Interlayer distance maps of the 
moir\'e unit cell of twisted bilayer graphene at 
$\theta \approx 3.1^\circ$, computed with (a)~PBE-D3, 
(b)~foundation MACE (OMAT-D3), and (c)~fine-tuned MACE. 
The foundation model systematically overestimates the 
interlayer separation and fails to reproduce the spatial 
contrast between stacking regions, while fine-tuned MACE 
recovers the PBE-D3 result with high fidelity.}
\label{fig:interlayer_validation_tblg}

\bigskip
\bigskip

\begin{center}
\includegraphics[width=0.9\textwidth]
{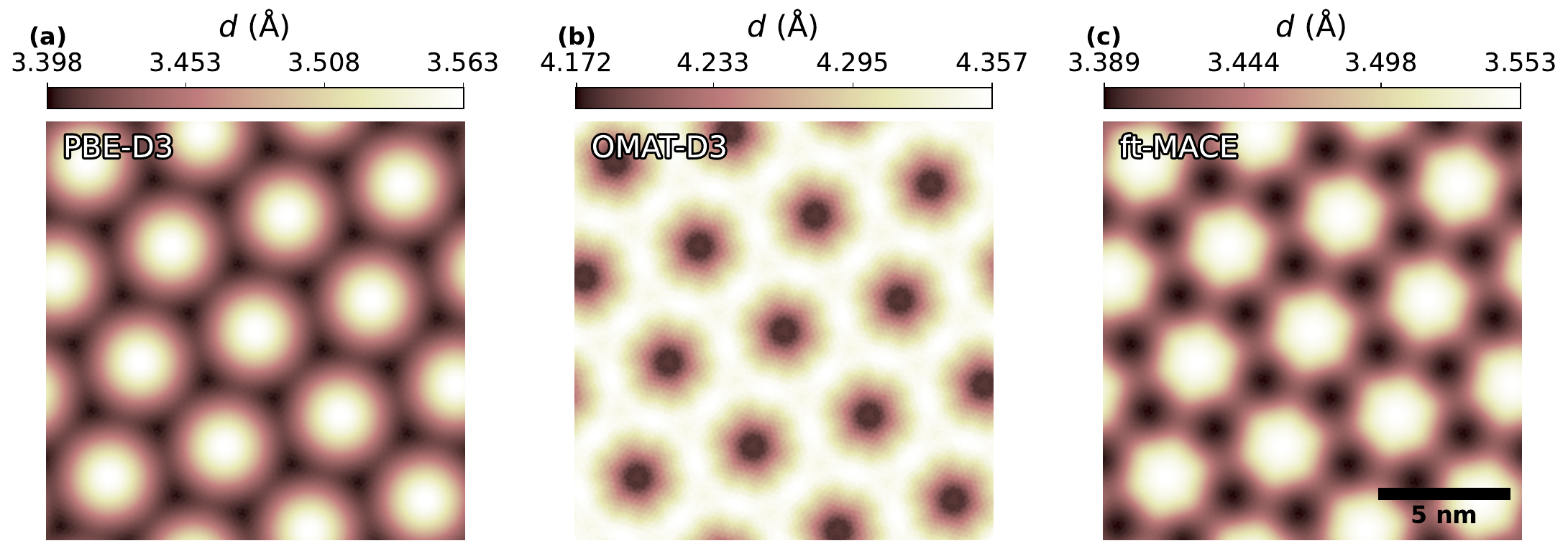}
\end{center}
\captionof{figure}{Same as 
Fig.~\ref{fig:interlayer_validation_tblg} for twisted 
bilayer \textit{h}-BN at $\theta \approx 3.1^\circ$.}
\label{fig:interlayer_validation_hbn}

\newpage

\begin{center}
\includegraphics[width=0.94\textwidth]
{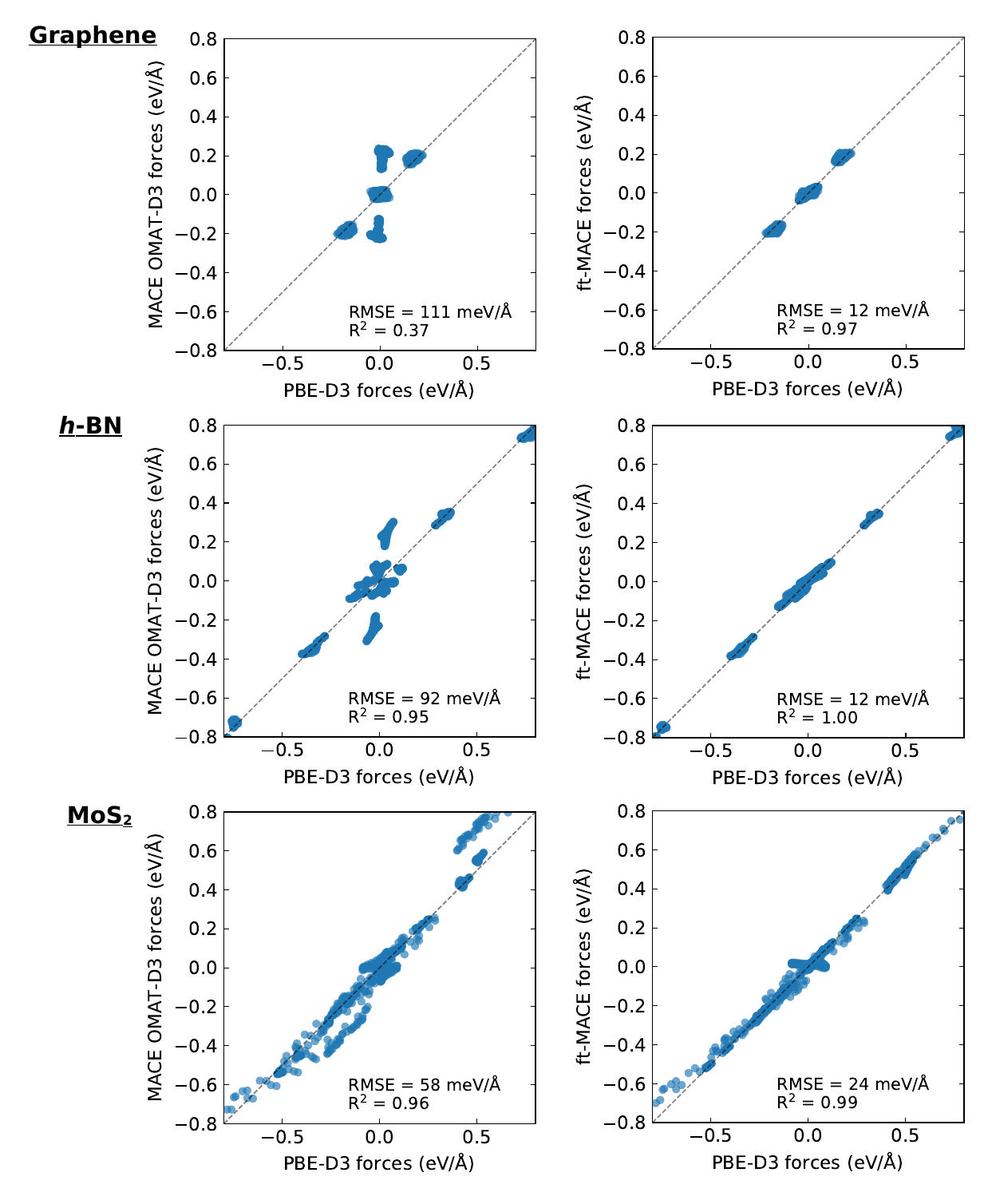}
\end{center}
\captionof{figure}{Force parity plots of foundation MACE (OMAT-D3, left column) and fine-tuned MACE (ft-MACE, right column) against PBE-D3 reference values, evaluated on quasi-one-dimensional moir\'e structures for graphene (top), \textit{h}-BN (middle), and MoS$_2$ (bottom), following the surrogate validation strategy of Ref.~\cite{georgaras2025accurate}. The improvement in force accuracy after fine-tuning confirms the reliability of the fine-tuned models for moir\'e simulations.}
\label{fig:quasi1D}

\bigskip
\bigskip

\noindent\textbf{B. Atomistic Modeling}: Phonon spectral function and unfolding procedure\\

\begin{center}
\includegraphics[width=0.8\textwidth]
{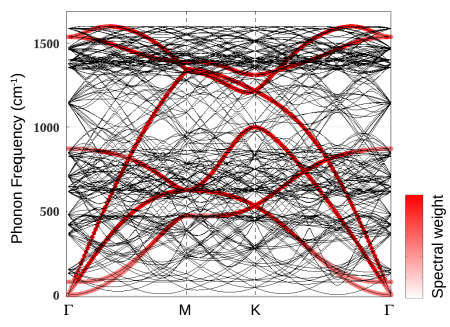}
\end{center}
\captionof{figure}{Illustration of the phonon unfolding procedure for a twisted bilayer graphene supercell. The dense folded branches of the moir\'e supercell (black) are projected onto the primitive-cell Brillouin zone, recovering a clean intensity-weighted dispersion (red).}
\label{fig:unfoldingscheme}

\newpage

\noindent\textbf{C. Reconstruction-Induced Strain Landscape}\\

\begin{center}
\includegraphics[width=\textwidth]
{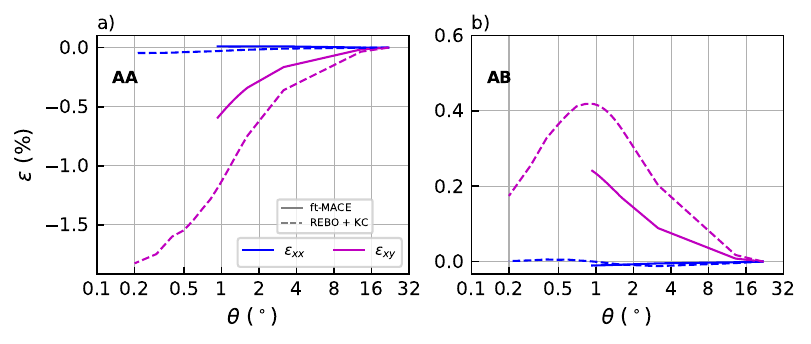}
\end{center}
\captionof{figure}{Twist-angle dependence of the 
$\varepsilon_{xx}$ and $\varepsilon_{xy}$ strain 
components in the AA and AB stacking regions of twisted 
bilayer graphene, predicted by ft-MACE and the REBO+KC 
classical force field. The REBO+KC potential 
systematically overestimates strain amplitudes by a 
factor of three to four relative to fine-tuned MACE over 
the full range of twist angles studied.}
\label{fig:strainAAABMACEnREBO}

\newpage

\begin{center}
\includegraphics[width=0.7\textwidth]
{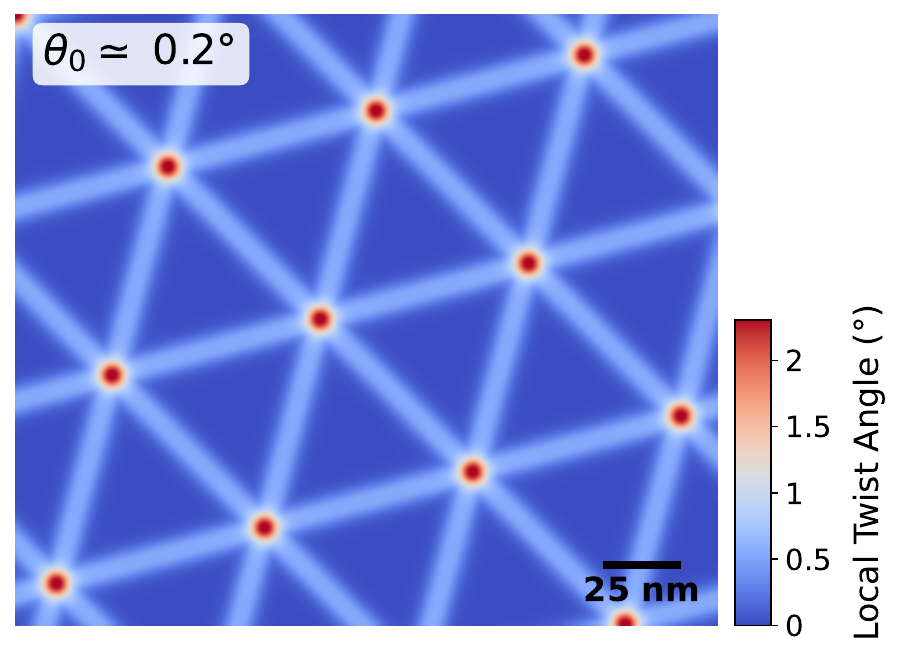}
\end{center}
\captionof{figure}{Spatial map of the local twist angle 
deviation $\Delta\theta(\mathbf{r})$ in small-angle 
twisted bilayer graphene at $\theta_0 \approx 0.2^\circ$, 
predicted by the REBO+KC classical force field. The map 
confirms the same qualitative pattern observed with 
fine-tuned MACE at accessible angles: AB domains exhibit 
$\Delta\theta < 0$ while AA nodes show $\Delta\theta > 0$, 
with the AB domain twist continuing to approach zero at 
this very small angle.}
\label{fig:localtwistangle_REBOKC}

\bigskip
\bigskip

\begin{center}
\includegraphics[width=\textwidth]
{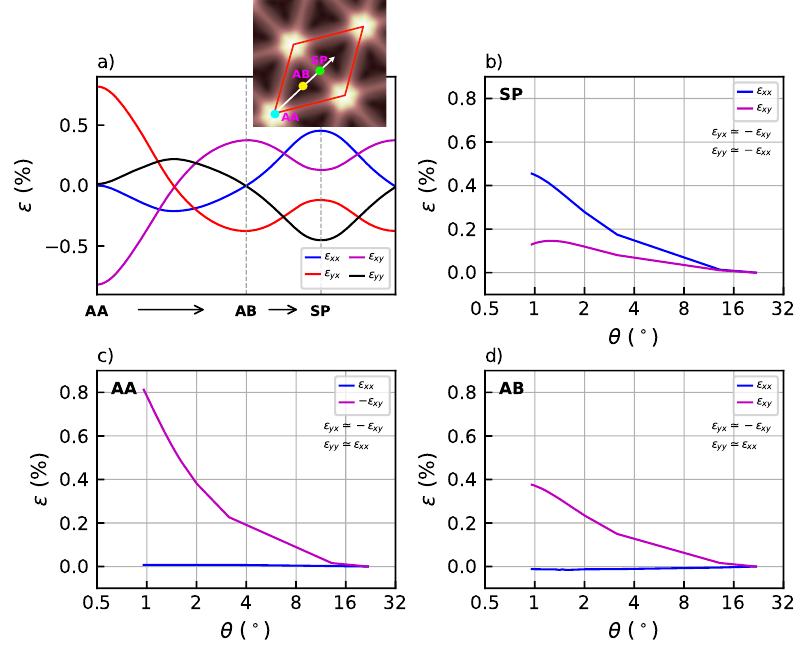}
\end{center}
\captionof{figure}{Local strain tensor components along 
the high-symmetry path at $\theta = 0.93^\circ$ (a) and 
twist-angle dependence of the strain in the SP, AA, and AB stacking regions (b--d) for twisted bilayer 
\textit{h}-BN in the parallel orientation.}
\label{fig:localstrain_hbn_a0}

\bigskip
\bigskip

\begin{center}
\includegraphics[width=\textwidth]
{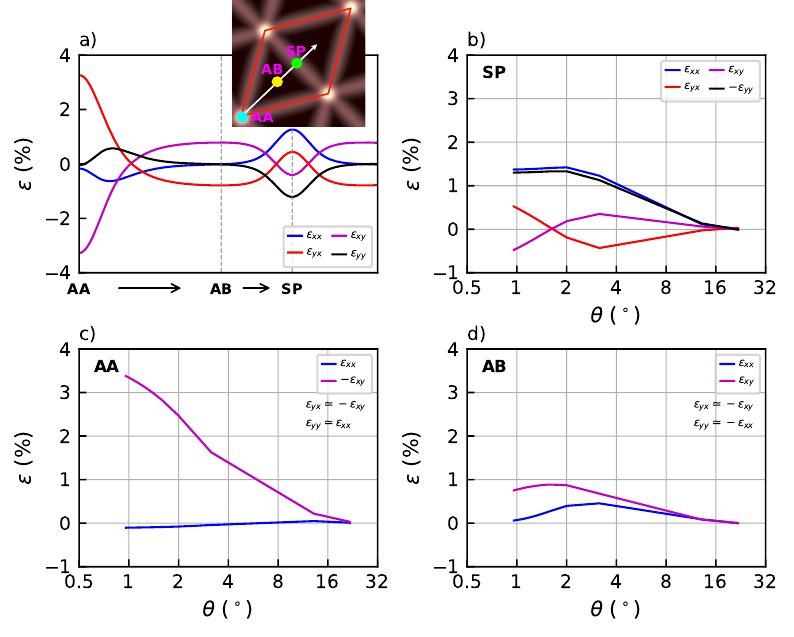}
\end{center}
\captionof{figure}{Local strain tensor components along the high-symmetry path at $\theta = 0.93^\circ$ (a) and twist-angle dependence of the strain in the SP, AA, and AB stacking regions (b--d) for twisted bilayer MoS$_2$ in the parallel orientation.}
\label{fig:localstrain_mos2_a0}

\bigskip
\bigskip

\begin{center}
\includegraphics[width=\textwidth]
{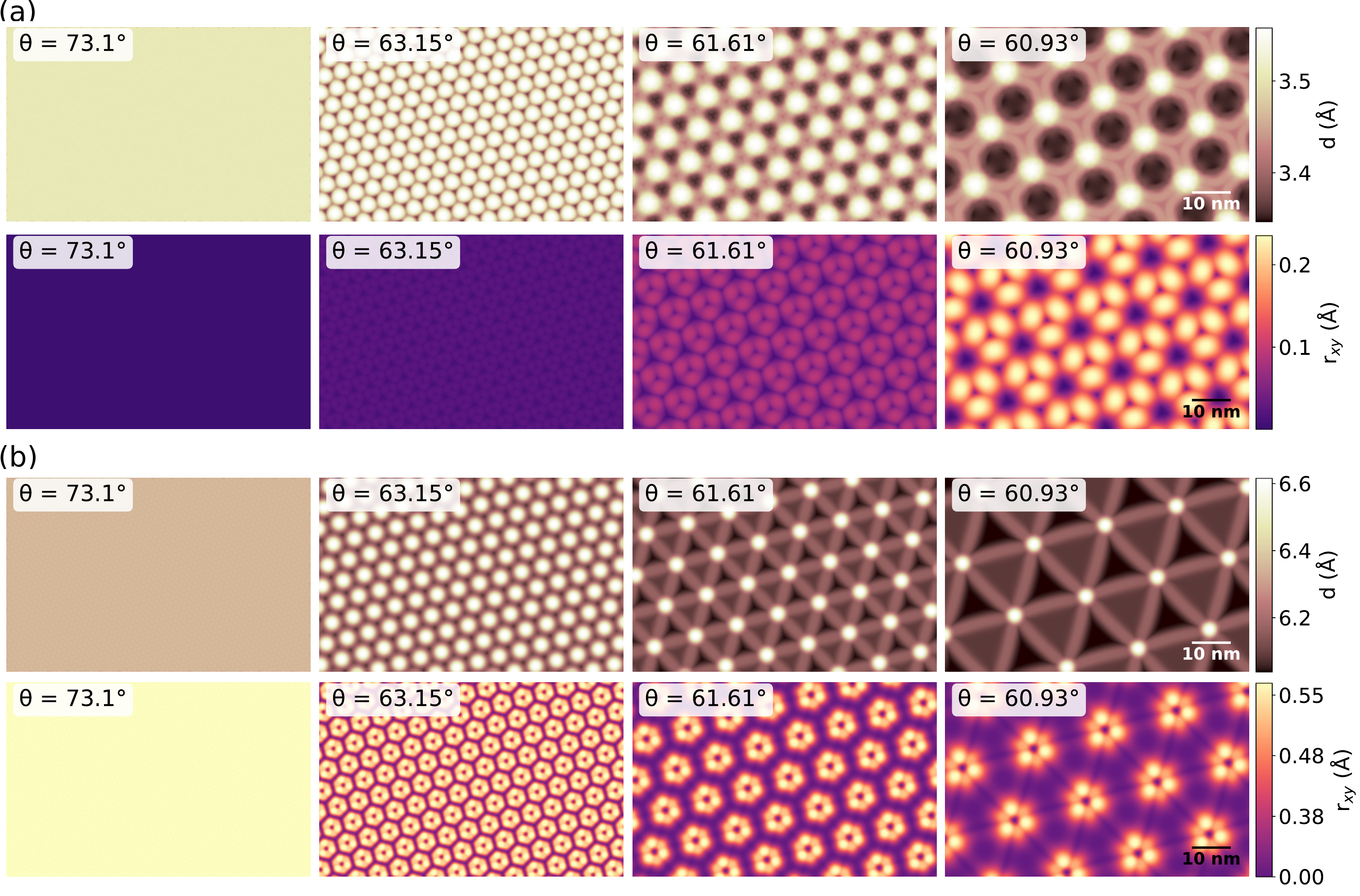}
\end{center}
\captionof{figure}{Interlayer distance and in-plane 
displacement magnitude maps of the moir\'e unit cell for 
twisted bilayer (a)~\textit{h}-BN and (b)~MoS$_2$ in 
the antiparallel orientation, at representative twist 
angles. The distinct contrast between AB$'$ and BA$'$ 
domains is more pronounced in \textit{h}-BN than in 
MoS$_2$, consistent with the difference in layer 
structure discussed in the main text.}
\label{fig:interandintralayer_a60}

\bigskip
\bigskip

\begin{center}
\includegraphics[width=\textwidth]
{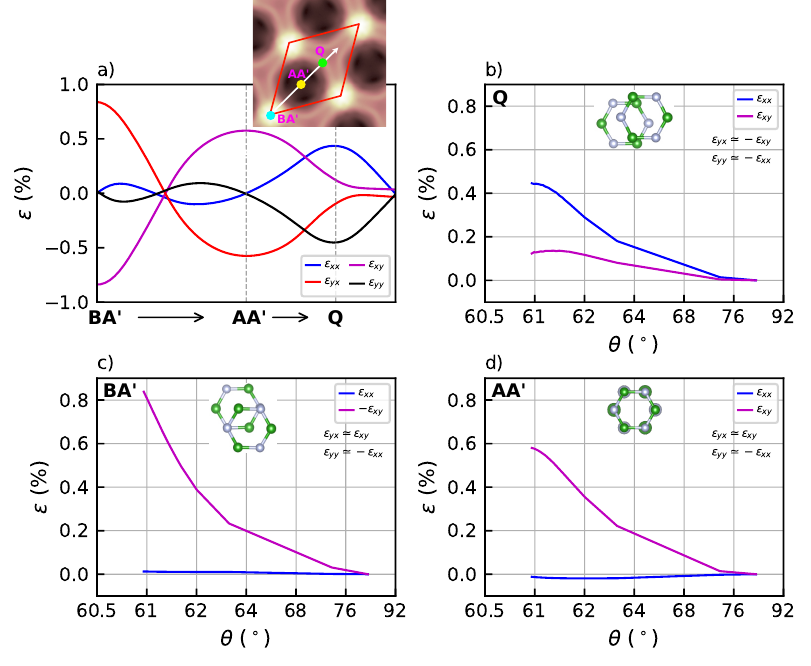}
\end{center}
\captionof{figure}{Local strain tensor components along 
the high-symmetry path at $\theta = 0.93^\circ$ (a) and 
twist-angle dependence of the strain in the Q, BA$'$, and AA$'$ stacking regions (b--d) for twisted bilayer 
\textit{h}-BN in the antiparallel orientation.}
\label{fig:localstrain_hbn_a60}

\bigskip
\bigskip

\begin{center}
\includegraphics[width=\textwidth]
{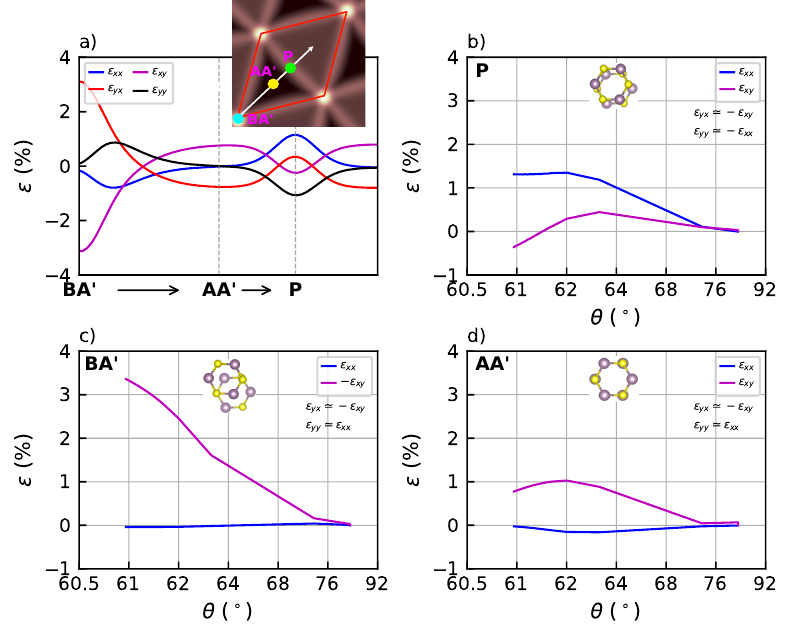}
\end{center}
\captionof{figure}{Local strain tensor components along 
the high-symmetry path at $\theta = 0.93^\circ$ (a) and 
twist-angle dependence of the strain in the P, BA$'$, and AA$'$ stacking regions (b--d) for twisted bilayer 
MoS$_2$ in the antiparallel orientation.}
\label{fig:localstrain_mos2_a60}

\end{document}